\documentclass[conference]{IEEEtran}
\input epsf
\usepackage{graphicx}
\usepackage{hyperref}
\usepackage[finalnew]{trackchanges} 
\addeditor{sc}
\addeditor{li}
\addeditor{fb}
\addeditor{mb}
\addeditor{cf}
\newcommand{\footurl}[1]{\footnote{\texttt{\url{#1}}}}

\newcommand{\visit}{\textsc{VisIt}}      \tcregister{\visit}{1}
\newcommand{\flash}{\textsc{Flash}}      \tcregister{\flash}{1}
\newcommand{\echo}{\textsc{Echo}}        \tcregister{\echo}{1}
\newcommand{\gadget}{\textsc{P-Gadget3}} \tcregister{\gadget}{1}
\newcommand{\enzo}{\textsc{Enzo}}        \tcregister{\enzo}{1}

\newcommand{\ray}{\texttt{[Ray~Casting: Compositing]}} \tcregister{\ray}{1}
\newcommand{\kb}{\texttt{[Kernel-based]}}              \tcregister{\kb}{1}
\newcommand{\tril}{\texttt{[Trilinear]}}               \tcregister{\tril}{1}
\newcommand{\rast}{\texttt{[Rasterization]}}           \tcregister{\rast}{1}

\newcommand{\yt}{\texttt{yt}}           \tcregister{\yt}{1}
\newcommand{\mpiforpy}{\texttt{mpi4py}} \tcregister{\mpiforpy}{1}
\newcommand{\numpy}{\texttt{NumPy}}     \tcregister{\numpy}{1}
\newcommand{\scipy}{\texttt{SciPy}}     \tcregister{\scipy}{1}

\newcommand{\XeonPhi}{Xeon Phi} 
\newcommand{\Xeon}{Xeon} 
\newcommand{\Intel}{Intel}
\tcregister{\Intel}{1}
\newcommand{\advixe}{Advisor}
\newcommand{\cmthr}{CoolMUC-3} \tcregister{\cmthr}{1}
\newcommand{\sng}{SuperMUC-NG} \tcregister{\sng}{1}
\newcommand{\phtwo}{SuperMUC Phase 2} \tcregister{\phtwo}{1}

\newcommand{\py}{{Python}}   \tcregister{\py}{1}
\newcommand{\cy}{{Cython}}   \tcregister{\cy}{1}
\newcommand{\mesa}{{Mesa}}   \tcregister{\mesa}{1}
\newcommand{\ospray}{OSPRay} \tcregister{\ospray}{1}

\tcregister{\XeonPhi}{1}
\tcregister{\Xeon}{1} 

\begin{document}
\title{Honing and proofing Astrophysical codes on the road to Exascale. Experiences from code modernization on many-core systems} 
\author{
  \authorblockN{
    Salvatore Cielo\authorrefmark{1},
    Luigi Iapichino\authorrefmark{1},
    Fabio Baruffa\authorrefmark{2}, 
    Matteo Bugli\authorrefmark{3} and
    Christoph Federrath\authorrefmark{4}
  }
  \authorblockA{
    \authorrefmark{1}
     Leibniz Supercomputing Centre of the Bavarian Academy of Sciences and Humanities\\ Garching b. M\"unchen, Germany - salvatore.cielo@lrz.de, luigi.iapichino@lrz.de
  }
  \authorblockA{
    \authorrefmark{2}
    Intel Corporation, Feldkirchen, Germany - fabio.baruffa@intel.com
  }
  \authorblockA{
    \authorrefmark{3} 
    {IRFU/Departement d’Astrophysique, CEA/DRF-CNRS-Universit\'e Paris Diderot, CEA-Saclay, France} - matteo.bugli@cea.fr
  }
  \authorblockA{
    \authorrefmark{4} Research School of Astronomy and Astrophysics, Australian National University \\ Canberra, ACT 2611, Australia - christoph.federrath@anu.edu.au
  }
}

\maketitle
\begin{abstract}
The complexity of modern and upcoming computing architectures poses severe challenges for code developers and application specialists, and forces them to expose the highest possible degree of parallelism, in order to make the best use of the available hardware. 
The Intel\textsuperscript{\textregistered}\ Xeon Phi\texttrademark\ of second generation (code-named Knights Landing, henceforth KNL) is the latest many-core system, which implements several interesting hardware features like for example a large number of cores per node (up to 72), the 512 bits-wide vector registers and the high-bandwidth memory.
The unique features of KNL make this platform a powerful testbed for modern HPC applications. The performance of codes on KNL is therefore a useful proxy of their readiness for future architectures. In this work we describe the lessons learnt during the optimisation of the widely used codes for computational astrophysics \gadget{}, \flash{} and \echo{}. Moreover, we present results for the visualisation and analysis tools \visit{} and \yt. These examples show that modern architectures benefit from code optimisation at different levels, even more than traditional multi-core systems. However, the level of modernisation of typical community codes still needs improvements, for them to fully utilise resources of novel architectures.

\end{abstract}
\IEEEoverridecommandlockouts
\IEEEpeerreviewmaketitle

\section{Introduction}\label{s:intro}

In order to keep pace with Moore's law, modern computing architectures are steadily growing in complexity. As a consequence, an increasing responsibility is imposed on developers and users of HPC applications: only exposing a high degree of parallelism one can exploit the available features offered by the hardware. More specific examples of this statement are the growing number of CPU nodes in high-end HPC systems, the number of cores in a node, the large vector registers of the cores, the memory hierarchy which requires streaming access to data for optimal performance. 
{Moreover, accelerated HPC systems pose the additional requirement of offloading the most computationally-intensive parts to their GPUs or accelerators, which always involves the users' attention or active work.}

Computational astrophysics is a research field where traditionally an extensive use of HPC resources is made. Although its vibrant scientific community has been constantly striving for the successful use of large-scale HPC facilities, also exploiting synergies with fields like Big Data and AI, the experience suggests that most of typical astrophysical applications consist of legacy code. For them, getting performance on current computing architectures can become problematic.

The optimization of simulation code, or the \textit{ab initio} development of modern applications have arisen to be an urgent task in the scientific community \cite{ColindeVerdiere2015,mlb16,mrk17,wkq17,sgc16,ipcc17,pwt18}. An underlying and often implicit point in this process is the identification of a target computing architecture for the optimization. In general, optimizations can be very machine-specific, like for instance the use of intrinsic instructions to get the most effective vectorization on a given architecture. On the other hand, other techniques aiming at merely exposing parallelism are more general and ensure portability of the resulting code.

A key requirement for a target architecture for optimization is to present hardware features that are interesting for the developer, either due to their being innovative, or because they are part of a trend which is going to stay for future products. With this in mind, in this work we want to consider the Intel\textsuperscript{\textregistered}\ Xeon Phi\texttrademark\ of second generation, code-named Knights Landing (henceforth KNL). This many-core processor implements a number of interesting solutions (a more detailed overview will be given in Section \ref{s:system}), potentially able of delivering performance for modern codes, if a high degree of parallelism at all levels and a cache-friendly access to memory is provided. Although the KNL product line has been recently discontinued \cite{xeonphi2018discontinued} by \Intel{}, we claim that its unique combination of hardware features still has a definite relevance for performance optimization on current and upcoming architectures.  

In this work we present the performance profiling on KNL of some broadly used software for astrophysical simulation and visualization. 
{More specifically, three code for astrophysical fluid mechanics} (\gadget{}, 
{\flash{} and \echo{}}) and two applications for data analysis and visualization (\visit{} and \yt{}) are part of our study. Although this list of applications covers only a small part of the use cases in computational astrophysics, our sample is however representative of codes with different levels of optimization, and can provide insights on the readiness of astrophysical code for future, pre-Exascale systems. 
{Our intention is not to cross-compare the work done on the single codes and their performance,
but rather to illustrate the properties of KNL through a number of real-life experiences based
on the chosen applications. As different codes have different strengths and weaknesses, we used quite a varied pool of evaluation tests, or \emph{Key Performance Indicators} (KPIs); in Table \ref{t-kpi} we show an exhaustive summary of those used for the different codes.  This process also tells much about the applications themselves, while helping to}
understand which architectural features, among the ones in KNL, are more useful for performance and necessary on future systems.

\begin{table*}[ht]
\caption{All Key Performance Indicators  (KPIs) presented for the considered codes. Unsatisfactory (-), irrelevant (0), satisfactory (+) and very satisfactory (++) results are indicated, together with relevant comments. For \yt{}, as baseline/optimized versions we mean Anaconda/\Intel{} \py{} (see text).}
\begin{footnotesize}
\begin{tabular}{c|cccccccccc}
\hline
{Code~\textbackslash{}~KPI} & 
\begin{tabular}[c]{@{}c@{}}Speedup \\ versus \\ baseline \end{tabular} &
\begin{tabular}[c]{@{}c@{}}Shared- \\ memory \\ parallelism \end{tabular} &
\begin{tabular}[c]{@{}c@{}}Speedup \\ from \\ SMT\end{tabular} & 
\begin{tabular}[c]{@{}c@{}}Energy \\ measure\end{tabular} & 
\begin{tabular}[c]{@{}c@{}}Node-level \\ scaling\end{tabular} & 
\begin{tabular}[c]{@{}c@{}}Large-scale \\ scaling\end{tabular} & 
\begin{tabular}[c]{@{}c@{}}Speedup \\ from \\ SIMD \end{tabular} &
\begin{tabular}[c]{@{}c@{}}MCDRAM \\ flat mode \end{tabular} & 
\begin{tabular}[c]{@{}c@{}}Comparison \\ with Xeon\end{tabular} &
\begin{tabular}[c]{@{}c@{}}Best HPC \\ method or \\ strategy \end{tabular} \\ 
\hline
\gadget{} & +   & +             &    & +  &     &    &              & 0          &     &                \\
\flash{}  & ++  & pure MPI      &    &    &     & +  & +            & unfeasible & -   &                \\
\echo{}   & +   & tested OpenMP & -  &    & +   &    & - , roofline & 0          &     &                \\ 
\hline                                                                
\visit{}  &     & pure MPI      & +  &    & +   & +  & no control   & unfeasible &     & Kernel-based   \\
\yt{}     & ++  & -             & -  &    & +   &    & no control   & 0          & +   & \cy{}         \\
\label{t-kpi}          
\end{tabular}
\end{footnotesize}
\end{table*}

The paper is structured as follows:
in Section \ref{s:system} we describe the main features of the KNL architecture and of the cluster where most of the presented tests are performed.
In Sections \ref{s:gadget} to \ref{s:yt} we present the tools used in this work, describing in particular their level of optimization, and discuss their performance results on KNL. Our conclusions are drawn in Section \ref{s:conclusion}.

\section{System and software stack} \label{s:system}
\subsection{KNL overview and investigated features}
\label{ss:KNL}
The Intel Xeon Phi of second generation (KNL) is a many-core processor designed for use in HPC systems.
In the following we summarize its main features and innovations; we focus on model 7210F, i.e. the one used for all our test runs (see Sections below), when not otherwise specified. We refer the reader elsewhere (e.g. \cite{knl17}) for a more detailed hardware description. 
\begin{itemize}
\item The KNL 
{can support a high degree of data parallelism, because} it has got up to $72$ cores ($64$ in the 7210F), many more than Intel Xeon\textsuperscript{\textregistered}\ processors of comparable launch time (for example, the ones of forth generation, code-named Broadwell, BDW), but with a low clock frequency ($1.3$\ GHz).
\item The KNL cores allow \emph{simultaneous multi-threading} (SMT) up to four threads per physical core, although in our experience using more than two brings seldom any performance benefit.
\item The cores feature 512-bits-wide vector registers, as {in the much newer} \Intel{}\textsuperscript{\textregistered}\ \Xeon{}\textsuperscript{\textregistered}\ Scalable Processor (code-named Skylake, SKX) and twice the size of those on BDW. The use of such registers is supported by the \Intel{}\textsuperscript{\textregistered}\ AVX-512 instruction set \cite{zhang16}. This feature is often also referred to as SIMD (Single Instruction, Multiple Data).
\item The KNL node has on-package high-bandwidth memory, based on the multi-channel dynamic random access memory (MCDRAM). On model 7210F, 16\ GB of MCDRAM with bandwidth $460$\ GB/s are available, a performance more than $5\times$ better than the system DDR4 RAM (larger, 96\ GB, but with bandwidth $80.8$\ GB/s; \cite{asai16b}).
\item 
{These two memory components} can be arranged in 
{two} different configurations,
known as memory modes
{ and} selected at boot time: the MCDRAM can be used 
as last-level cache (\textit{cache mode}), or the two 
can be exposed as separately addressable spaces (\textit{flat mode}; \cite{asai16b}).
\item The data locality in the access to cache is not trivial in a system with lots of cores. To ensure the lowest possible latency and the highest bandwidth, the KNL can be booted in so-called clustering modes, which determine the access path to memory controllers. We refer the reader to \cite{va16} on this topic; in the practical use, the \textit{quadrant} mode has emerged as the most widely used, while other clustering modes are profitable only for small groups of highly tuned applications.
\end{itemize}
{The most peculiar aspect is that the \emph{per core} computational power of the KNL is \emph{inferior} to the traditional multi-core architectures (\Xeon{}), in favor of a more exposed parallelism at all levels; a clear design choice of this product line, which later on has been partly reflected in some of the features of the \Xeon{} Scalable Processor.}

\subsection{Computing system and software environment}
\label{s:cluster}

{A list of all the machines and environments used in the paper is provided for quick reference in Table \ref{tab-envs}.}

Unless otherwise specified, we run on the \cmthr{} cluster\footurl{https://doku.lrz.de/display/PUBLIC/CoolMUC-3} (see Table \ref{tab-envs}) at the \emph{Leibniz Supercomputing Centre of the Bavarian Academy of Science and Humanities} (LRZ). The system consists of 148 Intel Xeon Phi 7210F nodes, connected to each other via an Intel\textsuperscript{\textregistered}\ Omni-Path high performance network in fat tree topology. 
Concerning software, from Intel Parallel Studio XE 2018 we used the Fortran and C compilers. Part of the analysis makes use of Intel\textsuperscript{\textregistered}\ Advisor 2019.
{Description and links for all other systems are provided in the text when needed.}

\begin{table*}[ht]
\caption{List of all environments and machines used, in order of introduction in the text. Unless otherwise specified, we refer to environment of \cmthr. The KNL architectures are used for our actual measurements, the others for perfomance comparison. }
\centering
\begin{tabular}{c|cccccc} 
\hline
Machine     & Architecture          & Model      & Owner & Physical cores & Threads per  & Purpose   \\
            & (codename, TLA)       &            &       & per node       & core (max)   &           \\ 
\hline
\cmthr      & Knights Landing, KNL  & 7210F      & LRZ   & 64             & 4            & Main platform\\
PCP cluster & Knights Landing, KNL  & 7250       & CINES & 68             & 4            & Energy measurements \\
\hline
\phtwo      & Haswell, HSW          & E5-2697 v3 & LRZ   & 28             & 2            & Performance comparison \\
\sng        & Skylake, SKX          & 8174       & LRZ   & 48             & 2            & Performance comparison\\
\label{tab-envs}
\end{tabular}
\end{table*}

\section{\gadget{}} \label{s:gadget}
\textsc{Gadget} \cite{springel2005cosmological} is a cosmological, fully hybrid parallelized (MPI plus OpenMP) TreePM, Smoothed Particle Hydrodynamics (SPH) code. In this scheme, both gas and dark matter are discretized by means of particles. 
Additional physics modules use different sub-resolution models, to properly treat processes which are far below the resolution limit in galaxy simulations. Our work is performed on a code version dubbed as \gadget{} \cite{bma2016sph}, based on the latest public release \textsc{Gadget-2} \cite{springel2005cosmological}\footurl{http://wwwmpa.mpa-garching.mpg.de/gadget/ }.

Some of the core parts of the code recently underwent a modernization, as reported by \cite{ipcc17}. In that work, a representative code kernel has been optimized by focusing on OpenMP concurrency, data layout and vectorization, resulting in a speedup up to $19.1\times$ on a KNL with respect to the un-optimized baseline code. 

\gadget{} has been successfully run at machine scale on large HPC systems, up to about 300 000 cores and $O(10^4)$ MPI tasks. Nonetheless the profiling of its shared memory parallelization showed serious bottlenecks \cite{ipcc17}. This performance problem has been solved in an isolated code kernel, but as of today the proposed solution has not been back-ported to the main code. From this viewpoint, \gadget{} is a typical simulation code, widely used in the community, with a relatively efficient MPI scaling, 
{although little effort was spent}
in the direction of node-level performance. 

The focus of our work on \gadget{} is twofold. First, we present an initial back-porting on the full code of the solutions developed for the kernel in \cite{ipcc17}, with a verification of that performance improvement (Section \ref{s:gadget-perf}). Secondly, we show a performance tuning of the energy to solution for a multi-node workload, based on the exploration of the parameter space at run-time (ratio of the number of MPI tasks and OpenMP threads per node; Section \ref{s:gadget-energy}).

\subsection{\gadget{}: Backporting and performance verification} \label{s:gadget-perf}
The work described in this section is a follow-up of the one described in \cite{ipcc17}. That paper focused on a representative kernel isolated from \gadget{}, on which a number of code improvements brought to a performance gain on both the target architecture of that study (the \Intel{} \Xeon{} Phi co-processor of first generation, code-named Knights Corner) and newer systems including KNL. Here we show that the performance gain obtained on the isolated kernel can be reproduced when said code improvements are back-ported into \gadget{}.

Some context of the back-porting development is necessary:
\begin{itemize}
\item among all optimizations introduced by \cite{ipcc17}, the back-porting is limited to the one concerning the threading parallelism, which removed most of the lock contention in the OpenMP part of the kernel;
\item for simplicity, the back-porting is limited only to the \texttt{subfind\_density} kernel in  \gadget{}, where the isolated kernel was extracted from.
\item the isolated kernel was OpenMP-only, whereas in the back-porting the OpenMP improvements were interfaced with the MPI section of the full code.
\end{itemize}
In order to ensure a consistent comparison, we tested the new full code on the same workload used for the isolated kernel: a cosmological simulation evolving $2 \times 64^3$ particles for a small number of time-steps.

The code has been run on a single KNL node using 64 cores in quadrant/flat configuration, with memory allocation on DDR, although unfortunately the code does not benefit from any specific cluster or memory mode.
For our comparison we used four MPI tasks, with 16 OpenMP threads each, a configuration we know as balanced and efficient from our previous work on KNL. The thread affinity has been set by the variables \texttt{OMP\_PROC\_BIND="spread"} and  \texttt{OMP\_PLACES="cores"}.

For the performance comparison we measure the timings of three different code versions. The first one, dubbed \textit{original}, consists of the baseline code in a “out-of-the-box” system environment with default machine configuration (Intel C++ Compiler 16.0, at the time of the tests) and no KNL-specific compilation flags. The second version, named \textit{tuned}, has the same source code as \textit{original} but implements a number of tuning solutions which do not require code modifications, namely the use of more modern compilers (Intel C++ Compiler 18.0) and of the compilation flag \texttt{-xMIC-AVX512} to target the highest vectorization ISA available on KNL. This category of tuning includes also an exploration of cluster and memory nodes but, as already mentioned, the code does not benefit from a specific combination.
The third code version, named \textit{optimized}, is developed on top of \textit{tuned}  and also includes the improvements in the threading parallelism in \texttt{subfind\_density}, described in \cite{ipcc17}.

\begin{table*}
\caption{Performance results of the \gadget{} tests in three different code versions, indicated in the first column. The table reports the time to solution and speedup with respect to the \textit{original} version for the whole application (second and third column, respectively) and for the \texttt{subfind\_density} function (fourth and fifth column. }
\centering
\begin{tabular}{c|cccc}
\hline
Code version & Time  & Speedup    & Time (fraction of total)  & Speedup \\
 & (total) [s]  &  (total) &  (\texttt{subfind\_density}) [s] &  (\texttt{subfind\_density}) \\ 
\hline 
\textit{original}  & 167.4 &  & 22.6 (13.5\%) &  \\
\textit{tuned} & 142.1 & 1.2$\times$ & 17.1 (12.1\%) & 1.3$\times$ \\
\textit{optimized} & 137.1 & 1.2$\times$ & 12.7 (9.3\%) & 1.8$\times$  \\
\label{t:gadget}
\end{tabular}
\end{table*}

The performance results of our tests on these three code versions for \gadget{} are summarized in Table \ref{t:gadget}. In the \textit{tuned} version the time to solution is improved by a factor of $1.2-1.3$. The effect of the optimization (third line in Table \ref{t:gadget}) adds no further speedup to the total timing of the code, because the optimized function \texttt{subfind\_density} takes a small fraction of the whole workload. However, this part of the code shows a speedup of $1.8\times$ in the version \textit{optimized} with respect to \textit{original}. This is larger than the factor $1.4\times$ we measured in the tests on the isolated kernel. The reason for it is that by decreasing the OpenMP thread spinning time, also the MPI imbalance is reduced, thus producing an additional positive effect on performance.
The back-porting of the threading optimization into \gadget{} exceeds the expected performance target and is therefore successful.

It is also interesting to compare the performance values on KNL with those measured on SuperMUC Phase 2\footurl{https://www.lrz.de/services/compute/supermuc/} 
{on a single node (} \Intel{} \Xeon{} E5-2697 v3, code-named Haswell, hereafter HSW; see Table \ref{tab-envs}). While the \textit{original} version runs $1.7\times$ slower (total time) on KNL,
the \textit{optimized} version is 
{just} $1.3\times$ slower{, numbers that decrease to $1.5\times$ (\textit{original}) and $1.16\times$ (\textit{optimized}) for the \texttt{subfind\_density} kernel by itself.} Many-core nodes 
profit from code optimization 
{better} than multi-core Xeon systems{, up to the point that} on optimized applications 
the performance gap between the two processor families tends to vanish.

\subsection{\gadget{}: Performance tuning and energy optimization} \label{s:gadget-energy}
It is well known that energy consumption and energy efficiency are primary concerns, as HPC systems evolve towards Exascale. While at run-time there are already solutions for optimizing the energy consumption of a job, like for example the energy-aware scheduling on SuperMUC Phase 2 at LRZ (Table \ref{tab-envs}; see also \cite{abb14} and the recent \cite{esc17}), it is also important to understand how the energy footprint of an application evolves while it is optimized. This question is especially interesting on KNL, because its energy efficiency has been claimed to be among its points of strength.

In this section we present the energy measurements done on \gadget{} during some steps of performance tuning, i.e.~performance improvements not requiring modifications of the source code. The work has been done on the \textit{optimized} version of \gadget{} (Section \ref{s:gadget-perf}), on a larger test case consisting on the evolution of a cosmological simulation of $2 \times 256^3$ particles, suitable for multi-node runs on KNL. 

We took our measurements on the pre-commercial procurement Bull Sequana X1000 KNL cluster at CINES\footurl{https://www.cines.fr/en/}, consisting of 168 Intel Xeon Phi 7250 nodes with 68 cores each @ $1.4$\ GHz. This system has been designed with a strong focus on energy efficiency. All nodes are in quadrant/flat configuration. On the system the Intel C++ compiler 17.0 was available. For the measurements we used the Bull Energy Optimizer (BEO) v.~1.0, a non intrusive energy profiler.

\begin{figure}
    \flushleft
    \includegraphics[width=.95\columnwidth]{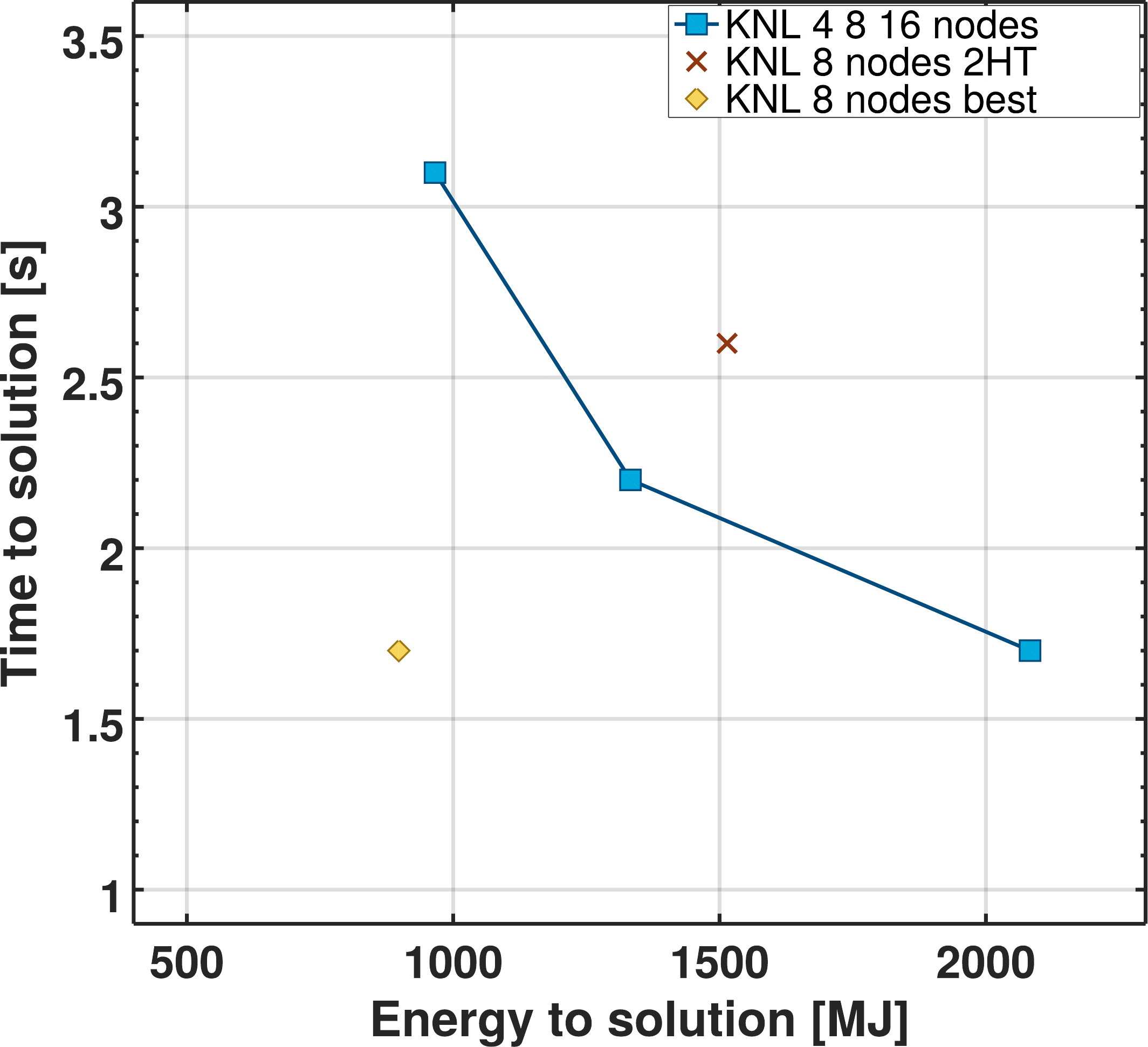}
    \centering
    \caption{Performance results (time to solution against energy to solution) for the multi-node energy measurements on \gadget{}. The points are labeled with the number of KNL nodes used for the respective runs; further indications are provided in the main text.} 
    \label{f:energy}
\end{figure}

When plotting time to solution against energy to solution as in Figure \ref{f:energy}, optimizing an application is equivalent to move the point on the plot towards the lower left corner.
The strong scaling of a code with ideal scaling should result in a vertical line descending the plot as the number of nodes increases. Otherwise, in the simplest case, one can assume that the energy footprint $E_{2N}$ of a job that is run on $2N$ nodes follows

\begin{equation}
    E_{2N} = E_N \times \frac{1}{P_{N \to 2N}}
    \label{e:energy}
\end{equation}

where $E_N$ is the energy consumption on $N$ nodes and $P_{N \to 2N} \in [0; 1]$ is the parallel efficiency going from $N$ to $2N$ nodes. One can verify that the blue squares in Figure \ref{f:energy} follow Equation \ref{e:energy}. For simplicity, we run a configuration of 4 MPI tasks per node and 16 OpenMP threads per task, leaving idle four cores per node. The red cross in Figure \ref{f:energy}, labeled as \textit{KNL 8 nodes 2HT} refers to a test with 8 KNL nodes and SMT with 2 threads per core (i.e. twice as many OpenMP threads per task as in the standard runs). The comparison with the run on 8 KNL in the standard configuration 
{(central blue square in Figure \ref{f:energy})} shows that hyper-threading is not a viable solution for optimizing the energy consumption for this workload.

The yellow diamond in Figure \ref{f:energy}, labeled as \textit{KNL 8 nodes best} is the outcome of a performance tuning. The parameter space of the ratio of MPI tasks to OpenMP threads per task has been explored on single-node tests, and the best combination in terms of smallest time to solution has been used for a full run. The optimal configuration consists of 32 MPI tasks per node and 4 OpenMP threads per task (here SMT is used) and therefore is very unbalanced towards a strong use of MPI. This can be understood with some previous knowledge of the code's behaviors: \gadget{} is known to have a well-performing inter-node communication scaling on large HPC systems, whereas the shared-memory parallelism in most code sections needs the full back-porting of the improvements described in Section \ref{s:gadget-perf} before being able to use efficiently $O(10-100)$ OpenMP threads. We stress that, for this code, this is a regime explored for the first time during the work on many-core systems.

{Thanks to} the performance tuning described above the time to solution improves by $1.5\times$ and the energy to solution by $1.3\times$.

{Although energy measurements have not been performed for other codes shown in this work, we claim that our two main findings (the link between energy consumption and parallel efficiency, and the role of performance tuning) have general interest and are worth being reported in this study on KNL.} Future work in the energy field will certainly move forward from performance tuning to code optimization, in particular exploring the performance of single code kernels. At even finer level of granularity, it would be interesting to evaluate the role played by single instructions like the vector instructions and their interaction with the clock frequency of the cores (e.g.~\cite{sib19}).

\section{\flash{}} \label{s:flash}

\flash{} \cite{fryxell2000FLASH} is a publicly available multiphysics, multiscale simulation code with a wide, international user base. Originally developed for modeling astrophysical thermonuclear flashes, it is appreciated for its multi-species nuclear physics capabilities, for the native adaptive mesh support (via the \textsc{Paramesh} package) and in general for the broad 
{range} of physical routines (including gravity, radiation treatment and \emph{magnetohydrodynamics}, MHD in the following), numerical methods and grid solvers it contains. These features made \flash{} the ideal choice for many CFD and physical applications. \flash{} is classically parallelized via pure MPI, though a hybrid scheme with OpenMP is included in the most recent release at the time of writing.

Its uses span the whole domain of astrophysics (and sometimes beyond\footurl{http://flash.uchicago.edu/site/publications/flash_pubs.shtml}), including --just to name a few-- 
interstellar turbulence and star formation \cite{fk12,fk13}
at very high resolution, but also the effects of \emph{Active Galactic Nuclei} on radiogalaxies \cite{antonuccio2010selfregulation} and galaxy clusters \cite{cielo2018reorienting}.

Recently, \flash{} has been optimized for reducing its memory and MPI communication footprints. In particular, non-critical operations are performed in single precision, when it is verified that this does not cause any impact on the accuracy of the results. This newly developed hybrid-precision version of \flash{} was able to run on SuperMUC Phase 2 (HSW nodes, Table \ref{tab-envs}) using less memory ($4.1\times$) and with a speed-up of $3.6\times$ with respect to the standard, double-precision version. The scaling up to machine scale on SuperMUC Phase 2 is also remarkable \cite{fki16}. 
{In the following, we refer to this version of \flash{} as the \emph{2016 version}, whose original optimizations were limited to hydrodynamic calculations, excluding the MHD part (see Table \ref{t:flash} for a summary).} 

\begin{table}[]
\centering
\caption{Internal references to the \flash{} version used in this work, documenting successive porting and optimizations. The optimizations were built on top of \flash{} 4.3.}
\begin{tabular}{c|c|c} \hline
\begin{tabular}[c]{@{}c@{}}FLASH version\\ (internal)\end{tabular} & Features & References \\ \hline
public version & \begin{tabular}[c]{@{}c@{}}double-precision\\ HD and MHD\end{tabular}                 & \cite{fryxell2000FLASH}       \\ \hline
2016 HD        & \begin{tabular}[c]{@{}c@{}}hybrid-precision HD,\\ general optimizations \end{tabular} & \cite{fki16}                  \\ \hline
2016 MHD       & \begin{tabular}[c]{@{}c@{}}2016 HD + \\ MHD from public version         \end{tabular} & this work                     \\ \hline
optimized MHD  & \begin{tabular}[c]{@{}c@{}}extension to MHD of \\ 2016 HD optimizations and \\ hybrid-precision scheme \end{tabular} & this work 
\label{t:flash}
\end{tabular}
\end{table}

In the present work we want to test whether such a hybrid-precision scheme can perform equally well also on KNL. In recent years, reducing the numerical precision of the computations has become a customary technique for gaining performance in fields like machine learning; its application in HPC to an architecture like KNL is therefore extremely interesting. 

\subsection{\flash{}: Weak Scaling} \label{ss:flash-ws}

We present an extension of the hybrid-precision version of \flash{} \cite{fki16} to include the treatment of MHD turbulence and investigate the effect of precision and vectorization optimization;
{we refer to this version of the code as \emph{optimized} (again, see Table \ref{t:flash})}.
The addition of MHD turbulence is an essential ingredient in star formation theories, however it introduces additional variables (the magnetic field components, for example) and equations which in turn increase memory, computation and communication requirements. For this reason it is important to re-assess scaling and optimization gains. 

\begin{figure}
  \flushleft
  \includegraphics[width=\columnwidth]{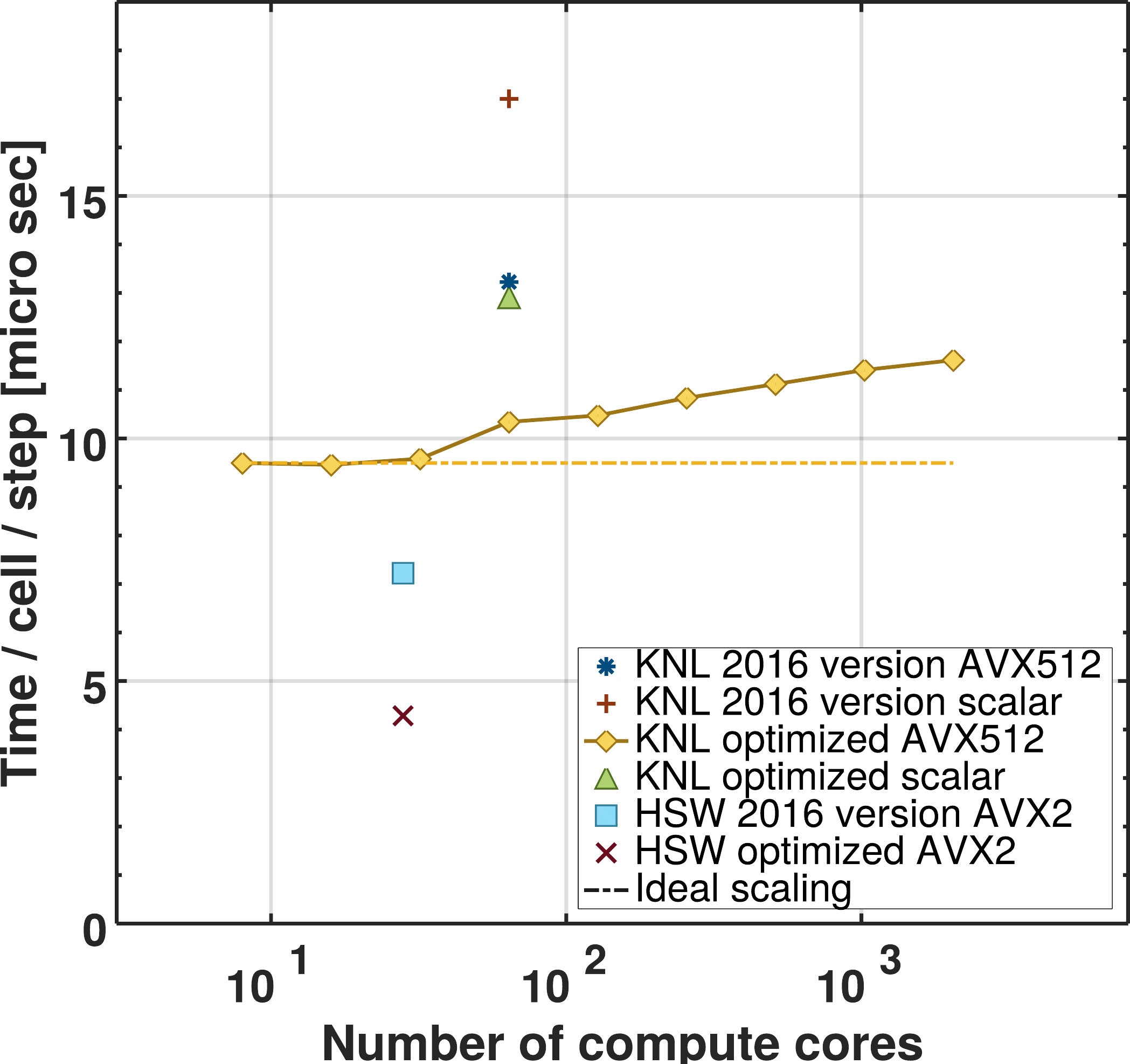}\vspace{0.3cm}
  \includegraphics[width=\columnwidth]{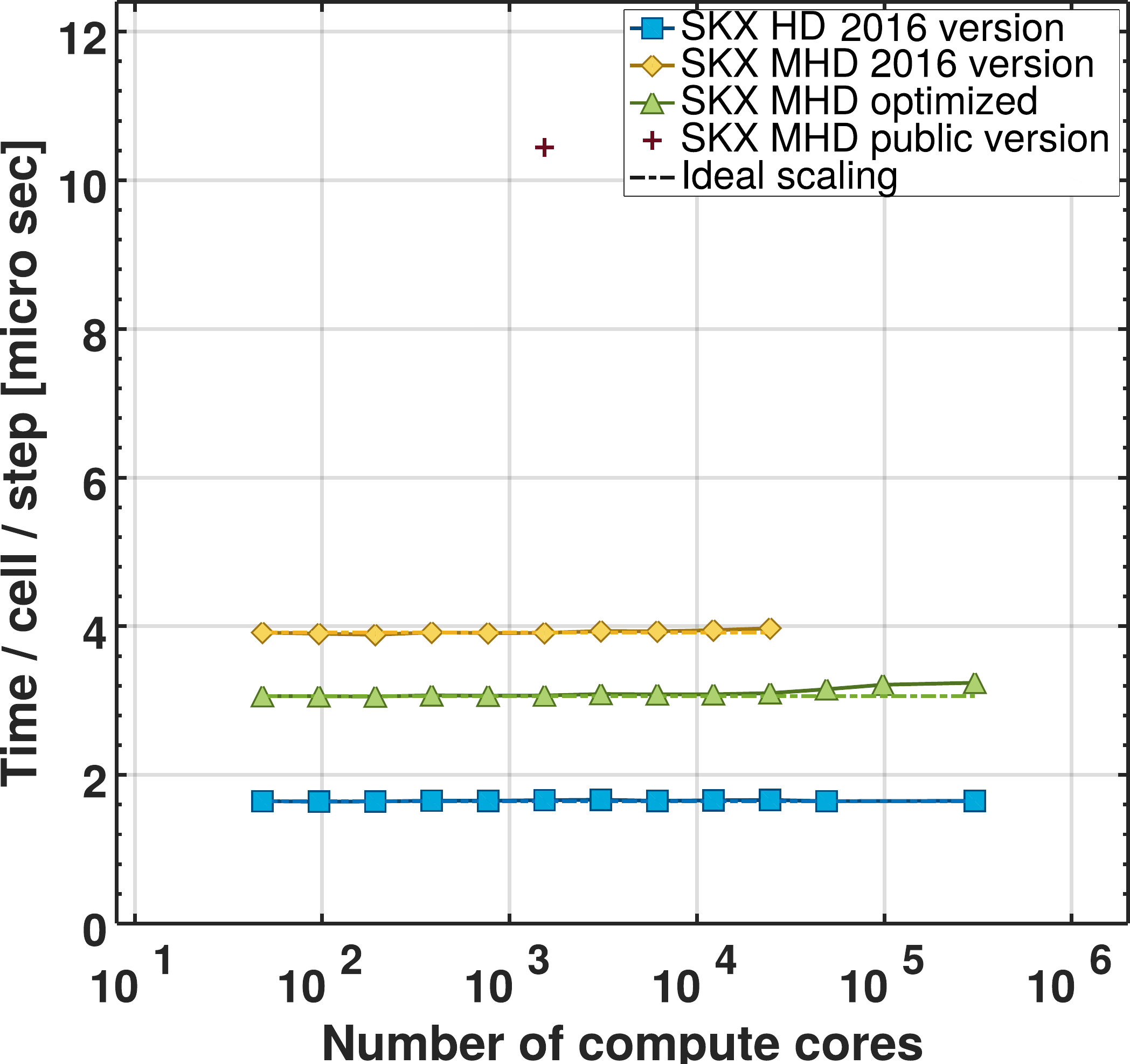}
  \centering
  \caption{Weak scaling tests with the \flash{} code on hydrodynamic (HD) and magnetohydrodynamic (MHD) turbulence. Top: scaling of the optimized MHD version (diamonds) on KNL. On a single node (64 cores) we show comparisons with the 2016 version of the pure HD runs. The star symbol is for vectorized code (AVX512), the triangle for the scalar (\texttt{-no-vec -no-simd}), and the plus sign for 2016 scalar. For comparison, 2016 and optimized versions on a single HSW node (square and cross, respectively) are also shown.
  Bottom: weak scaling for hybrid-precision, \texttt{-xCORE-AVX512} code on \sng{} (SKX system). The code benefits largely from vectorization and optimization (triangles), with respect to the 2016 MHD case (diamonds), though with the MHD overhead from the HD 2016 case (squares). The performance for public MHD version is shown as well (plus sign).}
  \label{f:flash_scaling}
\end{figure}

We thus perform a weak scaling test, i.e.~increasing the numerical resolution proportionally to number of cores used. The performance is then given by the time spent per time-step per unit computational cell, which remains constant in the ideal scaling case. We run the test up to 32 KNL nodes, in \emph{quadrant} cluster mode and \emph{cache} memory mode, because the memory allocation always exceeds the $16$~GB available in the MCDRAM. The results are displayed in 
{the top panel of}  Figure \ref{f:flash_scaling}{, where for comparison we show also the performance of the 2016 version, and the behavior of both on HSW architecture (see Table \ref{tab-envs}).}

{Concerning the effect of the optimizations,} the 
{MHD optimizations (diamond symbols) grant} a performance gain of a factor $1.28\times$ respect to the 
{2016 version} (star symbol,
computed on a full KNL node). A speedup of $1.25\times$ is instead obtained through AVX-512 vectorization, when compared to the scalar version (compiled with \texttt{-no-vec -no-simd}; triangle in the same panel). Among all tested codes, \flash{} has the largest speed-up due to vectorization. These two optimizations work in close synergy, as single-precision operations offer an additional factor of two on vector registers (compared with 2016 scalar version, plus symbol).

The scaling of the 
{optimized} version remains nearly ideal up to 32 cores, but degrades steadily when using more (diamonds in the top panel of Figure \ref{f:flash_scaling}). This behavior is not intrinsic of the application, but is rather caused by some system property. We can see this from the optimal scaling measurement obtained on \sng{}\footurl{https://doku.lrz.de/display/PUBLIC/SuperMUC-NG} (SKX nodes, Table \ref{tab-envs}), presented in the bottom panel of Figure \ref{f:flash_scaling}. 
{The triangles refer to the optimized MHD code, which outperforms the 2016 version (diamonds) also on SKX, and at odds with KNL shows acceptable scaling, up to the whole \sng{} system (304,128 cores).}
Up to this number of processors, 
{the MHD override still influences a bit scaling and performance ($2.35\times$ slower), when compared to non-MHD 2016 version (blue squares).}

Since both systems 
{are equipped with} the same interconnect (\Intel{} Omni-Path with a maximum bandwidth of $100$~Gbit/s per port), the difference we measure does not depend on the connection hardware.
As further evidence of this, different large-scale tests performed with other applications 
{in this work}
do not suffer from the same issue (e.g. \visit{}, cf. Section \ref{ss:visit-results}). These
results suggest that the library in charge of the parallel communication (\Intel{}\textsuperscript{\textregistered}\ MPI 2017, the default on \cmthr{}) is to be held responsible, also given the additional messaging load introduced by the MHD variables. 
The problem 
{might} be mitigated by introducing the OpenMP 
{layer} of parallelization within single KNL nodes, as implemented in the most recent version on \flash{}, or simply by updating the MPI version (e.g., version 2019, as for the \sng{} run). In any case, the scaling issue
{of \flash{} do not seem to occur on any of the tested Xeon systems. A deeper analysis of this feature is left for future work.}

When comparing the performance with the one achieved on other systems, we notice that the Xeon architecture profits from our optimization more than KNL. Indeed the 
{optimized} version on HSW (cross symbol in the upper panel of Figure \ref{f:flash_scaling}) is $1.69\times$ faster than the 2016 one (cross symbol); larger than the $1.28\times$ factor of the KNL). On a node-to-node comparison, the execution on HSW is $2.4\times$ faster than on KNL. As expected, the performance on SKX (diamonds and triangles in lower panel of Figure \ref{f:flash_scaling}
) exceeds the one on KNL by about $3\times$.

{To sum up, using hybrid precision on \flash{} is certainly a suitable solution to improve the performance of the code, although the reported test case shows that this technique is slightly more profitable on traditional Xeons than on many-core systems.}

\section{\echo{}} \label{s:echo}
\echo{} \cite{delzanna2007} is a finite differences, Godunov type, shock-capturing astrophysical code designed to solve the equations of magneto-hydrodynamic in general relativity (GRMHD). In the recent years it has been employed in numerous studies for modeling a wide range of high-energy sources, such as pulsar wind nebulae \cite{olmi2015}, neutron stars \cite{pili2014,pili2017} and thick accretion disks surrounding black holes \cite{bugli2018a}. 

On top of the original GRMHD scheme, the code currently includes 
{additional physical} effects such as magnetic diffusion and mean-field dynamo action \cite{bucciantini2013}, 
expressed in a covariant formalism consistent with the relativistic framework{. These features, that come with additional computational costs, have been applied to} the study of $\alpha\Omega$-dynamos in magnetized tori around black holes \cite{bugli2014} and relativistic magnetic reconnection \cite{delzanna2016}. 

The version 
{of \echo{}} used in this work presents a recently implemented hybrid MPI-OpenMP scheme \cite{bugli2018b,bugli2018c} that has vastly improved the performance of the original code{:} the multidimensional domain decomposition has allowed the efficient exploitation of large HPC facilities by running on up to 4096 nodes and 65536 MPI tasks \cite{resch2017}, 
{and} the additional layer of multi-threading introduced in the most computationally intensive fractions of the code has led to a significant increase in its scaling capabilities,
{ although it did not reduce the memory footprint significantly}. 
{The target system of the work just described was SuperMUC Phase 2, consisting of HSW nodes}. In the following we will focus mostly on 
{evaluating} the performance of the code on KNL, so to assess to what extent the 
{optimized} version of \echo{} can exploit the 
{architectural features of the Intel Xeon Phi. From this viewpoint, our approach to the optimization of \echo{}  is similar to the one for \gadget{}, namely we evaluate on KNL the performance of solutions originally developed on Xeon systems.}   

\subsection{\echo{}: Node-level Scalability}\label{ss:echo-scale}

\begin{figure}
  \centering
  \includegraphics[width=\columnwidth]{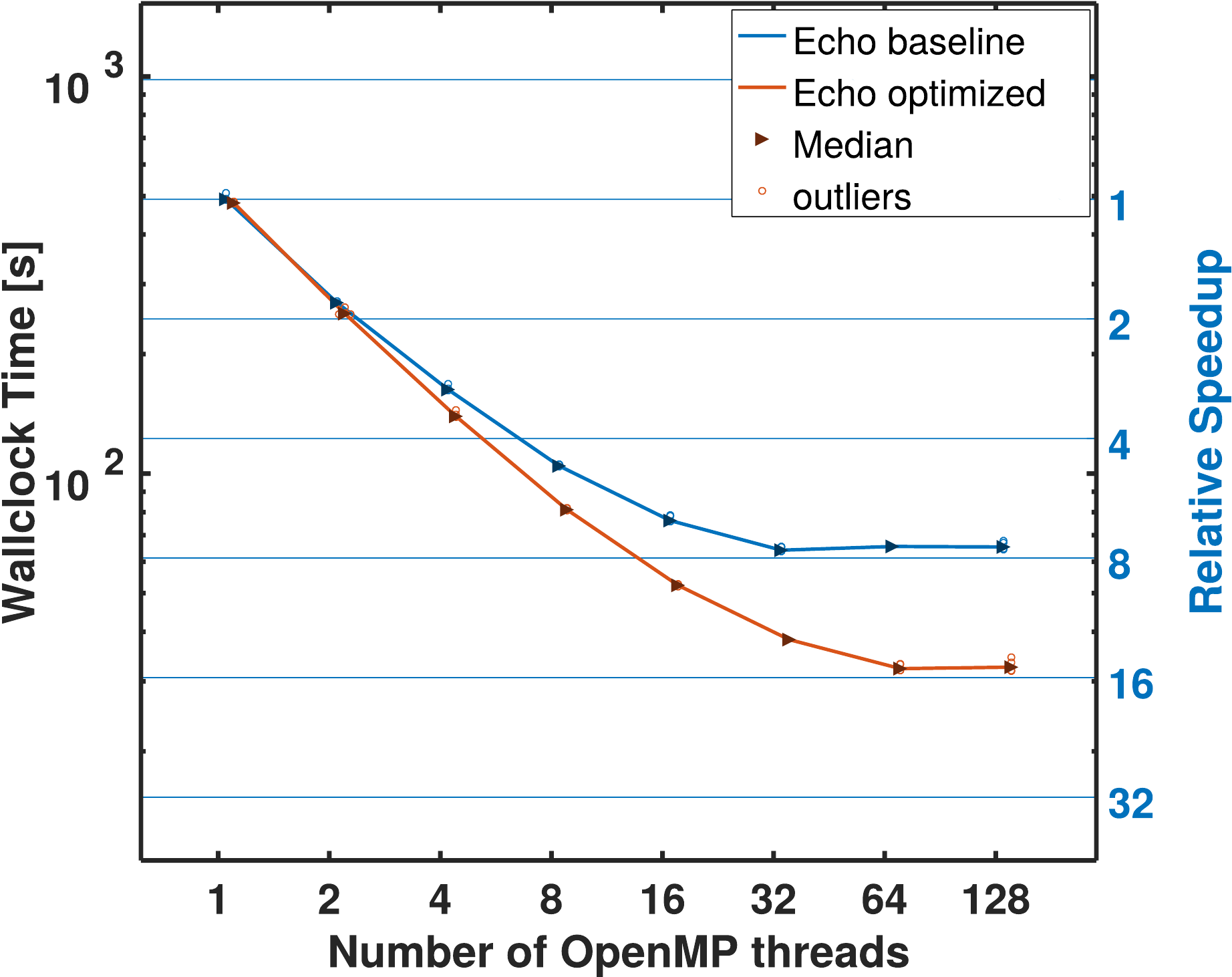}
  \caption{Single-node scalability test, showing wallclock execution time, for the baseline (red) and optimized (blue) versions of the \echo{} code. Parallel speedup, relative to the vector and optimized version, is readable on the $y$-axis, on the 
  {right-hand side}. 
  {The points mark the median of 20 measures. The statistical distribution around each point is very narrow, so the only (barely) distinguishable features are the occasional outliers (small empty circles).} Performance and scalability are both improved by a factor of 
  {2 at and beyond 64 threads}. }\label{f:echo_scaling}
\end{figure}

Figure \ref{f:echo_scaling} shows the results of a scalability test on 
{a single} KNL node on \cmthr{}. This pure-OpenMP version of the code 
runs a 
{5}-timestep evolution of a standard MHD test, i.e. the propagation of an Alfv\'en wave, travelling diagonally with respect to the coordinate axis. We adopt a domain size of $100^3$ cells (top panel),
{ and} run our test up to 64 threads (without SMT) and 128 threads (SMT with two threads per core). 
{Threads are always pinned to individual cores (we set \texttt{I\_MPI\_PIN\_DOMAIN=core} and \texttt{I\_MPI\_PIN\_ORDER=compact}), although the results are mostly insensitive to the pinning scheme, as long as threads are not allowed to migrate and we do not exceed one thread per core.}

The chosen performance metric is the wallclock-time to solution, displayed for both base (red line in Figure \ref{f:echo_scaling}) and optimized (blue line) version. The relative parallel speedup can be read on the $y$-axis as well. The points mark the median of 20 measures; a \emph{boxplot} statistical analysis is performed for each point, however we display those features only when clearly distinguishable, not to overcrowd the plot. For the same reason, each dataset is slightly jittered along the $x$-axis.
In this case all distributions are fairly narrow; only the occasional outliers (small empty circles) can (barely) stand out; this consistent performance is sign of system stability, and a suitable pinning scheme. 

The performance is improved by a factor up to $2\times$, 
{and} the code scales better, taking advantage of the full 64 cores on the KNL. The point at 128 threads shows that hyper-threading gives no advantage, when not degrading the performance altogether.

The improvement
of the optimized version on the KNL is in line with what observed on 
{HSW} \cite{bugli2018c}, with the additional bonus of larger scalability. 

This satisfactory result is also corroborated by the performance comparison with the time to solution on a single HSW socket (14 threads on SuperMUC Phase 2): on KNL with 64 threads the baseline version is a factor of $1.4$ slower than on HSW, but on the optimized version the gap reduces to $1.15$, showing as in Section \ref{s:gadget-perf} and in \cite{ipcc17} that the many-core architecture profits more than multi-core Xeons from code optimizations.

\subsection{\echo{}: Vectorization and Roofline Analysis}\label{ss:echo-roof} 
\begin{figure*}
  \centering
  \includegraphics[width=1.0\textwidth]{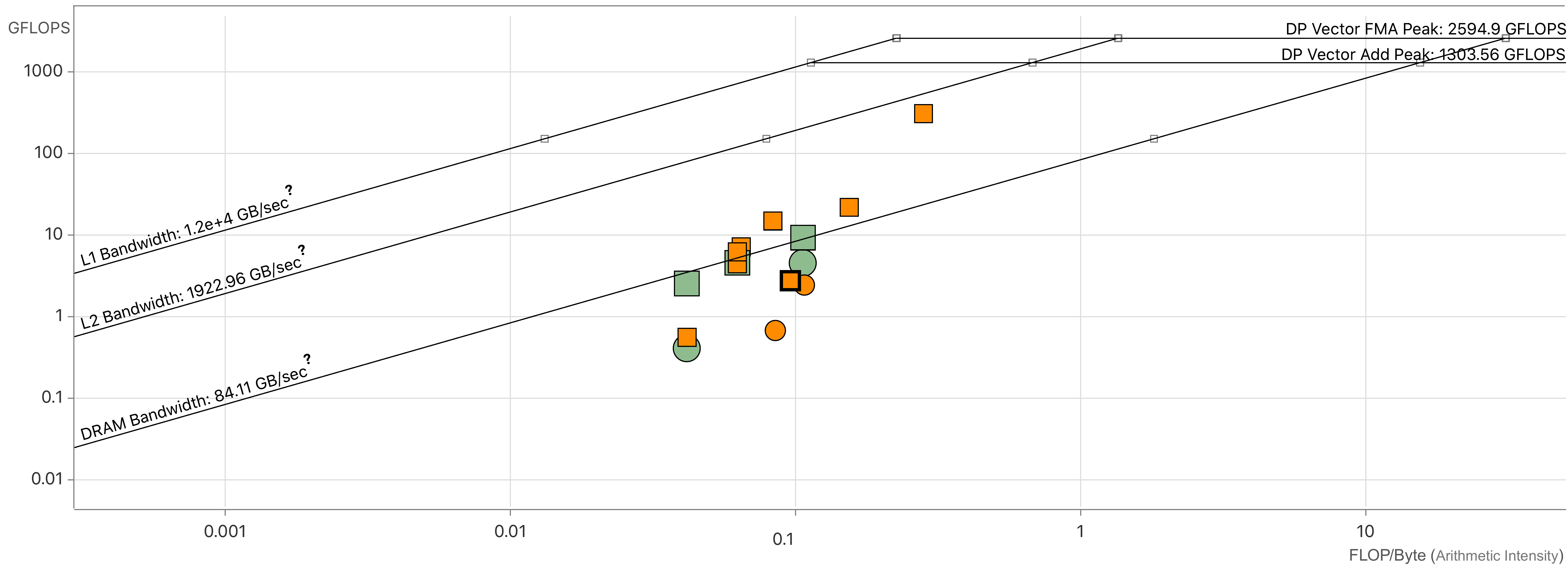}
  \caption{Roofline analysis, performed with \Intel{} \advixe{} 2019, showing the performance of \echo{} on \cmthr{}. Only the 
  {most time-consuming loops and functions} are shown. Squares: optimized version. Circles: baseline version. Coloring: elapsed time in percent, where green is larger than $13\%$ and orange is between $5\%$ and $13\%$ of the total execution time.
  }\label{f:echo_roofline}
\end{figure*}

{A remaining point to be addressed for the node-level performance of \echo{} is the potential gain due to vectorization on the large SIMD registers of KNL}.
To estimate the performance gain 
{due to vectorization,} we compare the execution times 
{of the optimized code, compiled with \texttt{-xMIC-AVX512}, with its scalar version compiled with \texttt{-no-vec -no-simd}}. The performance gain from vectorization is a modest factor of 
{1.18}, largely independent of threads number and optimization, and not higher than the one observed on 
{HSW}, despite the latter is using only AVX2 ISA on smaller vector registers. The explanation for this behavior is mainly related to the inefficient access to data in memory. This can be explored more closely by making use of the \textit{roofline analysis}, as we do in Figure \ref{f:echo_roofline}{, where we show the five most time-consuming loops and functions in \echo{} from tests on 64 threads. }
 
The roofline model for studying the performance of an HPC application has been first introduced by \cite{wwp09}. In brief, an element of a code (loop, function, or even the whole code) can be represented as a point into a plot of the performance in FLOPS on the $y$-axis against the arithmetic intensity 
{(i.e. the number of operations executed per byte read from memory, expressed in FLOP/B)} on the $x$-axis. The code element is classified as memory bound (left-hand side of the plot) if it is located under the diagonal “roof” defined by memory bandwidth of the system, or as compute bound (right-hand side of the plot) if it is under the horizontal line defined by the peak performance of the system. Figure \ref {f:echo_roofline} has been computed by making use of the so-called cache-aware roofline model \cite{ips14}: in this variant, the diagonal roofs represent the bandwidth of the different memory levels, as seen from the cores.
{As an important consequence}, the optimizations in the code that do not involve algorithmic changes are visualized as vertical shifts in the roofline plot towards higher performance.

The main message in Figure \ref{f:echo_roofline} is in fact the memory-bound nature of the code, 
{as the points constantly hit the DRAM bandwidth roof. 
The details of the algorithmic role of the single loops and functions represented in the figure go far beyond the scope of this discussion. However, it appears clearly that a significant improvement was achieved in the optimized version over the baseline, as squares appear above the correspondent circles of same arithmetic intensity, and closer to, or even above, the actual DRAM roof.} 

We recall from \cite{bugli2018c} that the main optimization between \echo{}'s baseline and optimized version is on the threading parallelism. Now that the application is mainly DRAM-bandwidth bound, the next step will be to modernize the access to data in memory to overcome this bottleneck. We anticipate that, to some extent, this step will allow an easier data movement to vector registers and thus a larger vectorization efficiency.
 
We finally tested whether this bandwidth-limited code could benefit from using the MCDRAM of the KNL in \emph{flat mode} (see Section \ref{s:system}). 
{Although it was possible to allocate all memory objects onto MCDRAM,} since the memory footprint of our runs is lower than 16~GB, we measured no statistically significant discrepancy with the allocation on DRAM or with the cache mode.

\section{\visit{}}  \label{s:visit}

\visit{} \cite{childs2012visit} is a parallel data processing and visualization tool broadly used in the astrophysics community and in many other numerical fields. 
Thanks to its several I/O plugins, it is commonly used on a very 
{broad} spectrum of simulations (e.g.~all of the above simulation codes are supported) as well as real data (e.g., FITS files, also as multidimensional data cubes). Its main strengths lie in the versatile (2D and) 3D rendering capabilities -- the ones we investigate in this work -- but most reduction and post-processing operations (selection/filtering, integration, time series analysis and a broad range of queries) can be performed as well, through \visit{}'s interactive GUI, or its well-integrated Python interface. The latter allows to invoke a batch scripting version of \visit{}, which is also the easiest option for use in HPC environment\footnote{Interactive options, both in GUI and in CLI mode, are possible, but of course require interactive access to the servers}. The main advantage of this workflow is to use the same facilities (computing nodes, storage, user environment, queuing system) for simulation productions and processing.

Despite 
the fact that many traditional scientific visualization tasks are 
executed on 
{accelerators like for instance the Graphical Processing Units} (GPUs), and optimized for those, modern scientific visualization tools can
{also} benefit from the massive parallelism offered by multi-core and many-core CPU-only 
{systems} at and above node level, as well as from 
{the large vector registers of modern cores}. \visit{} is no exception; besides the default GPU-accelerated rendering, at compile time one can 
{instead} 
{select} support by the \mesa{} 3D graphics library\footurl{https://mesa3d.org}. This choice is best suited for servers without displays or GL environment, such as the compute nodes on \cmthr{}, where the use of \visit{}'s \texttt{GUI} and \texttt{Viewer Window} is not possible. 

{Unlike previous architectures, the KNL was presented as a platform
    for graphical and CPU-intensive applications, coming from a co-processor
    lineage from the Intel collection. Taking on a task traditionally reserved
    to GPUs is a hard test for a CPU-like hardware, but in our study we present 
    an exhaustive exploration showing that some methods indeed reach a level of
    scaling and performance worthy of optimized HPC applications. 
    Later architectures such as SKX are also capable of similar tasks; 
    however such a comparison with a later generation would not be fair.}
More advanced options however exist (see Section \ref{ss:visit-discuss}). 
{The version of \visit{} we use for all our runs is 2.13.2, compiled with Mesa.}

\subsection{\visit{}: Ray-casting Methods and Scalability} \label{ss:visit-results}
{The key} aspect we investigate here is the choice of the visualization technique most suitable to a many-core architecture, among the many different ones offered by \visit{}. For our visualization study, we focus on \emph{ray casting} techniques, as they present several advantages over other types of visualizations (texturing, rasterization, etc.). These include easy scalability to large datasets, low-level compatibility with a wide range of data types (volumes, polygons, non-polygonal meshes, particles) and easy introduction of shading and after-effects \cite{wald2017ospray}. 

For \visit{} we perform two scaling tests: one over 32 nodes\footnote{The largest job allowed by the scheduler on \cmthr{}.}, comparing different ray-casting methods, and 
{another} on 
{a single} node, focusing on node-level optimization.

We always have \visit{} perform the same task, with each of the different 
{sampling} options available within the \ray{} method: \tril{}, \kb{} and \rast{}. 
{These algorithms have been especially developed for the use on supercomputers }\cite{childs2006raycasting} and adopt a multi-step scheme to parallelize on both domain and image elements.

We use a linear ramp as transfer function, and a sampling rate of 800 per ray, to produce a ${(1024\,\mathrm{pixel})}^2 $ image. With the chosen graphic options, the final image is 
{nearly independent of} the method we choose{, thus proving that the resolution of the rendered image is high enough to contain little trace of the sampling strategy}.

\begin{figure}
    \centering
    \includegraphics[width=\columnwidth]{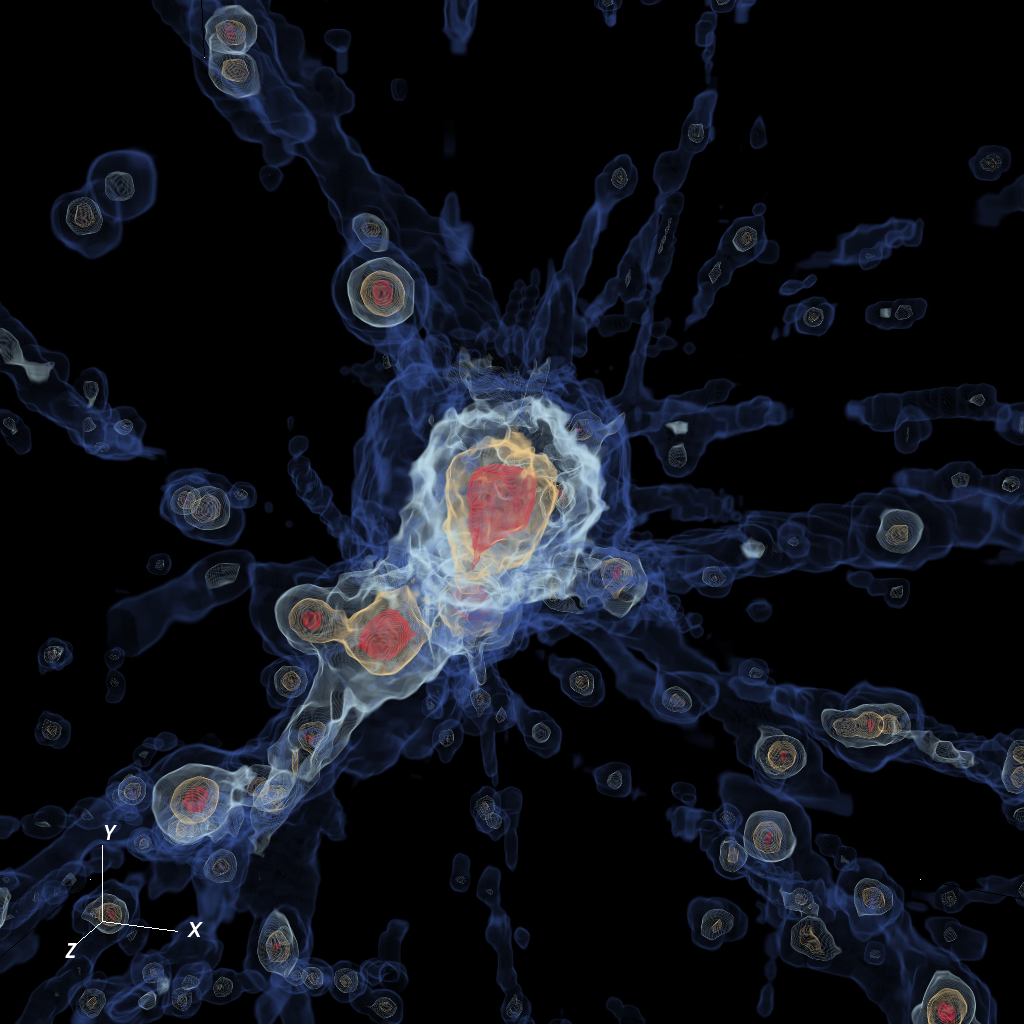}\vspace{10pt}    \includegraphics[width=\columnwidth]{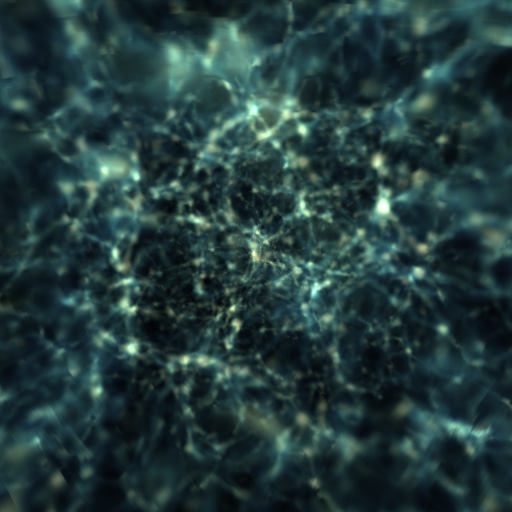}
    \caption{Volume rendering of the gas density, from a cosmological simulation run with the AMR code \textsc{Enzo}. This is the dataset used for the tests on \visit{} (zooming on the central $30$\ Mpc, top panel) and \yt{} (using the whole dataset with a size of $380$\ Mpc on a side, bottom panel). In the bottom panel one can notice that the resolution is better in the central region of the computational domain, because of the nested grid structure (so-called zoom simulation).
    } \label{f:visit_visual}
\end{figure}

The chosen dataset is a snapshot at redshift $z=0.05$ from a cosmological simulation \cite{ifk17} performed with the grid-based, adaptive mesh refinement (AMR) code \enzo{} \cite{enzo14}. A gas density 
{volume rendering is} shown in Figure \ref{f:visit_visual} (top panel). The size of the dataset is $2.4$\ GB, and the output consists of 6722 AMR grids and $1.9 \times 10^7$ mesh elements.

\begin{figure}    
    \centering
    \includegraphics[width=\columnwidth]{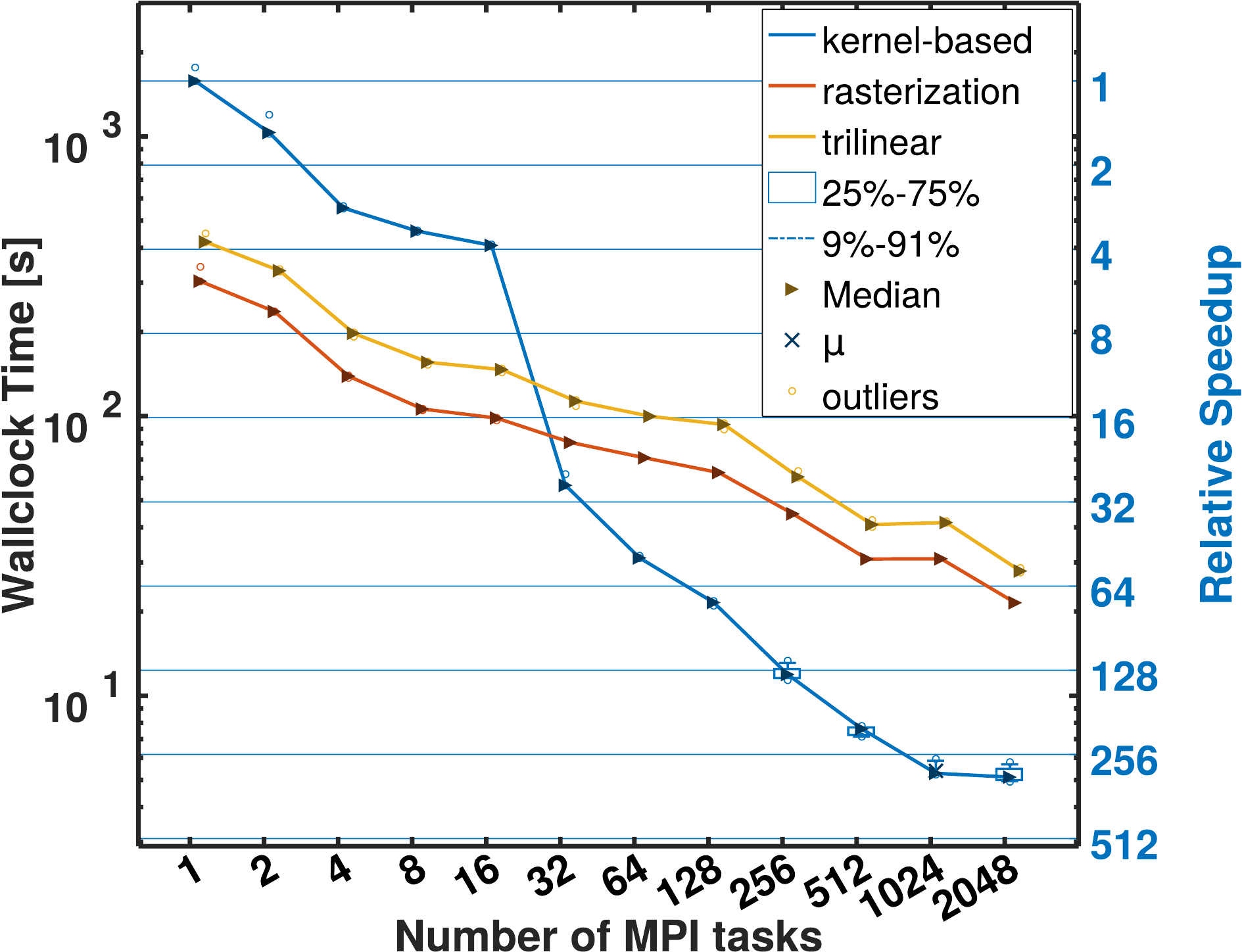}
    \caption{Wallclock execution time for \visit{} volume rendering scalability test, up to 32 nodes. Axes and errorbars have the same meaning as in Figure \ref{f:echo_scaling}, though 10 measurements per point were taken. The different lines compare the different techniques offered by the \ray{} renderer. 
    The \kb{} method, though the serial slowest, shows the best overall scaling.
}\label{f:visit_method}
\end{figure}

The plot in Figure \ref{f:visit_method} shows the scaling results of the rendering tests. As in Figure \ref{f:echo_scaling}, we perform a statistical analysis for each point (we took 10 measurements for each). This time the distributions are not as narrow, so when needed we also display the mean $\mu$, the 25th and 75th percentiles as box, the 9th and 91st percentiles as errorbars; outliers are still marked by empty circles, and the relative speedup is still shown on the right-hand side.

The times shown refer to the sum of the two main phases of the \visit{} rendering process: the \texttt{Pipeline} (i.e. the pre-processing to create a 3D scene) and the \texttt{Scalable Rendering} phase (in which a 2D image is created from the 3D elements).
The \kb{} method, though the serial slowest, shows large improvement past 16 cores, and the best overall scaling. The reason for this sudden speedup are still under investigation, but are found in the \emph{Scalable Rendering} phase. No other method achieves a similar parallelization degree in this step{; meaning that this method can gain the most in the image and domain decomposition}. This behavior is also reflected in the \kb{} line being the steepest of the plot, past 16 tasks. 

All tests hit a plateau around 1024 or 2048 tasks (16 or 32 nodes). Even though larger errorbars appear for the \kb{} method past 2 nodes, the performance is still by far the best. The second-fastest option is the \rast{}, although the scalability is utterly similar in \rast{} and \tril{}, and the difference lies just in the normalization.
{Our findings confirm that the strong scalability obtained by the developers up to 400 cores} \cite{childs2006raycasting} still holds on the KNL nodes.

\begin{figure}    
    \centering
    \includegraphics[width=\columnwidth]{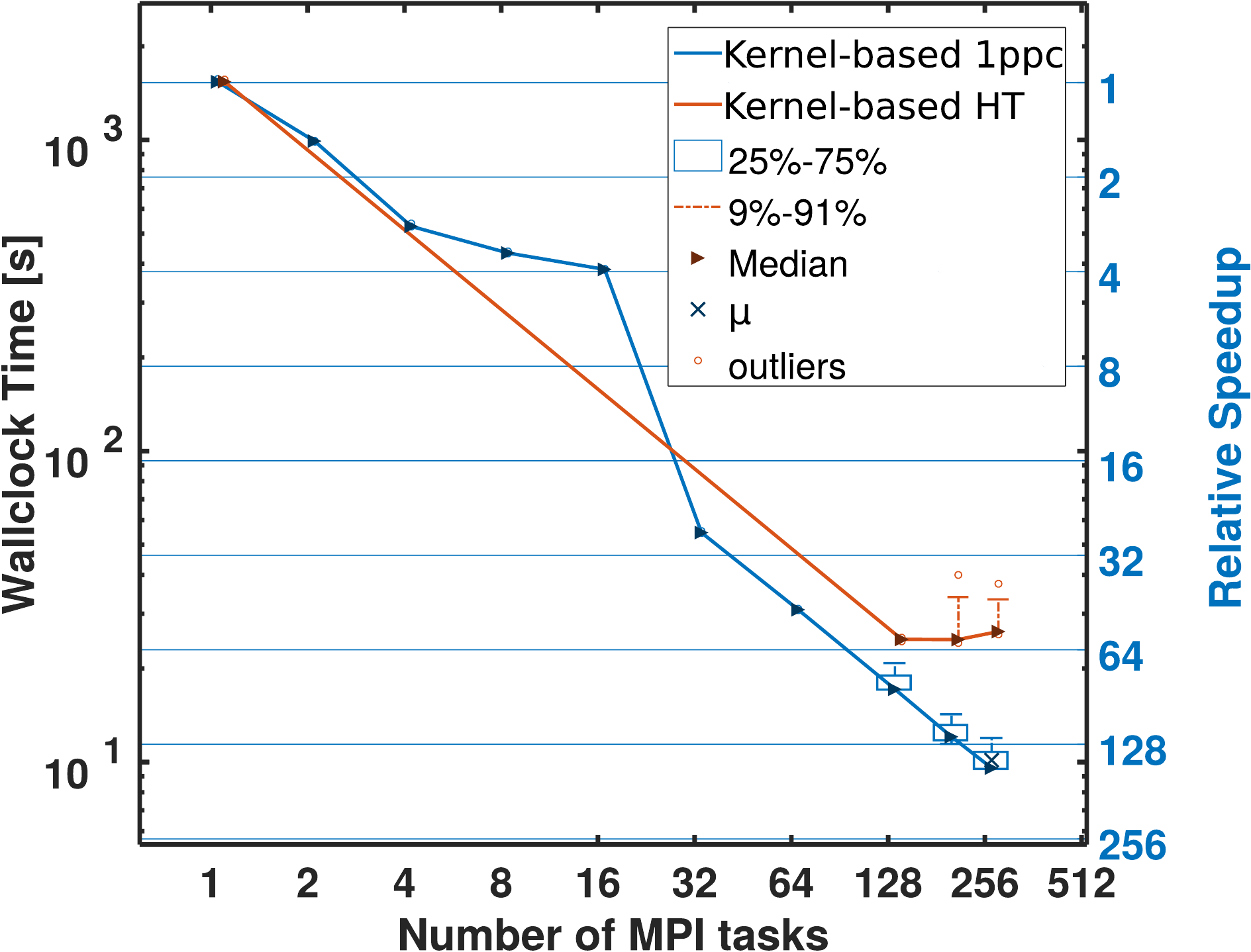}
    \caption{Wallclock execution time for \visit{} node-level performance test. Axes and errorbars have the same meaning as in Figure \ref{f:visit_method}, The method is the \kb{} for both lines. The \emph{HT} line (red) refers to hyper-threading, with 2, 3 and 4 threads per core. The \emph{1ppc} line (blue) uses instead 2, 3 or 4 nodes with one process per core each, for direct comparison. Two threads per core allow for the best speedup. }\label{f:visit_node}
\end{figure}

Finally, in Figure \ref{f:visit_node} we show the results of our node-level performance test.
We first investigate the effects of hyper-threading. The 
{orange} line in the figure contains three points at 128, 192 and 256 MPI tasks, obtained on a single node using 2, 3 and 4 threads per physical core (as well as a serial point, for reference). We observe that using 2 threads per core is an improvement over one, reaching a speedup of almost exactly 64, proof of efficient use of all the resources of the node. At 3 processes per core, the median performance is still stable around 64, even though with worse statistics, while at 4 the median execution time increases slightly.
We conclude that using two processes per core is the most efficient way to use a single KNL node for a \visit{} job, although spreading the same processes on more nodes (64 each), gives better speedup (comparison of the yellow and blue lines). In the latter case, inter-node communication is responsible for the larger errorbars we observe. 

\subsection{\visit{}: Discussion on CPU-defined performance} \label{ss:visit-discuss}
Our experience with \mesa{} \visit{} is overall satisfactory, as it shows good parallel speedup up to a few thousands of MPI tasks (which may increase for larger tasks). 
For comparison, profiling of \mesa{} \visit{} on earlier generations of \Intel{} \Xeon{} processors gave much poorer results when compared with GPU rendering (\cite{wald2017ospray}), even though in a different context. In that 
{paper those} authors present their work on the \emph{\ospray{} Distributed Ray-Tracing Framework} \cite{atipa201Xospray}, which can be integrated in \visit{} as essentially a parallel rendering engine, targeting non-accelerated HPC machines. \ospray{} installations have shown massive improvements over GL or \mesa{}, especially for real-time interactive visualization and when performing large tasks on KNL and \Intel{} \Xeon{} architecture \cite{wu2018visitospray}. 
While this is an appealing solution, its investigation will be left to future studies \cite{cielo2019supercomputing}.
For the moment we conclude that 
{the \mesa{} version of} \visit{} can 
{already} take a huge advantage from the KNL architecture, although this is also the kind of system where advanced solutions like \ospray{} can shine. Either option seems viable 
{on upcoming, pre-Exascale HPC systems}. Running some tool-assisted performance analysis is desirable in this perspective, but such investigation is hindered by the complex client-server relation between \visit{}'s main process and the engines it launches autonomously.

\section{\yt{}} \label{s:yt}
\yt{} \cite{turk2011yt} is an open-source, community-driven \py{} 
toolkit for analyzing and visualizing 3D data, in the form of meshes or particle sampling. Its compatibility with most broadly used simulation tools and the constant development by the active user community has proven very valuable in fields such as astrophysics, earth sciences, and molecular dynamics. \yt{}'s greatest strength is its deep integration with professional analysis tools used in observational astrophysics, available within the same \py{} environment, so that in many case the numerical and observational/experimental communities can share the same tools, such as data processing (often instrument-specific) pipelines, as well as math libraries and statistics packages.

{Although no specific optimization was applied on \yt{}'s source code previous to our work,
here we test several modernization strategies suitable for \py{} packages, aiming
to have \yt{} take better advantage of the KNL possibilities.}

{At variance with \visit{}, }
\yt{} is a \py{} package, thus its performance may depend on the distribution on top of which it runs. In this sense, it is representative of a broad and popular class of tools. 

For these reasons, \yt{} is a suitable testbed for the \Intel{}\textsuperscript{\textregistered}\ Distribution for \py{}$^*$\footurl{https://software.intel.com/en-us/distribution-for-python} (\Intel{} \py{} for brevity in the following). This distribution provides optimized versions of most of the libraries \yt{} depends on (e.g.~\numpy{}, \scipy{}, \mpiforpy{}) which yield substantial, documented performance improvements \cite{intel201Xpythonbenchmarks} on HPC systems, including many-core ones. 


{A large fraction of \yt{} users utilize the provided self-installation script, which builds an Anaconda \py{} distribution. However, using \Intel{}\textsuperscript{\textregistered} proprietary (though freeware) Distribution for \py{}$^*$\footurl{https://software.intel.com/en-us/distribution-for-python} (\Intel{} \py{} for brevity in the following) on Intel hardware is a strongly advisable synergy, on which Intel has spent significant resources over the last few years} (see \cite{intel201Xpythonbenchmarks}). 
{We thus begin our work on \yt{} with a comparison between the two \py{} distributions, to quantify the gains in performance or scalability that most users take the risk to miss out.}
To this aim, two equivalent Conda environments are created, based on \py{} 3.6.2 provided respectively by \Intel{} and Anaconda, where \yt{} 3.3.5 has been installed. 

\yt{} can in principle make use of several parallelization strategies: a message-passing scheme via the \mpiforpy{} package, a built-in OpenMP integration via \cy{} \cite{behnel2010cython}, or a hybrid combination of the two \cite{yt201Xparallel}. 
{Besides these possibilities, the tool offers support for embarrassingly parallel analysis of time series of datasets. In the following, we will review examples of these different strategies in use.}

\subsection{\yt{}: parallel projections with \mpiforpy{}}\label{ss:yt-projection}

In our first series of tests we measure the wallclock time of a \yt{} projection of gas density along the $x$-axis. According to \cite{turk2011yt}, the parallelization in the projection in \yt{} is managed by distributing the AMR grids of the dataset among the MPI tasks. The dataset is the same we used for the \ray{} rendering with \visit{} (cf. Section \ref{ss:visit-results} and Figure \ref{f:visit_visual}). The tests are performed on a single KNL node of \cmthr{}. Different cluster and memory configurations have been tested but showed no performance difference; the presented measurements refer to averages of 15 runs performed in the quadrant/flat KNL configuration. MPI thread pinning is beneficial on many-core systems, thus it was set by the environment variable \texttt{I\_MPI\_PIN\_PROCESSOR\_LIST =} $n$\texttt{-1}, where $n$ is the number of MPI tasks.

\begin{figure}    
    \centering
    \includegraphics[width=\columnwidth]{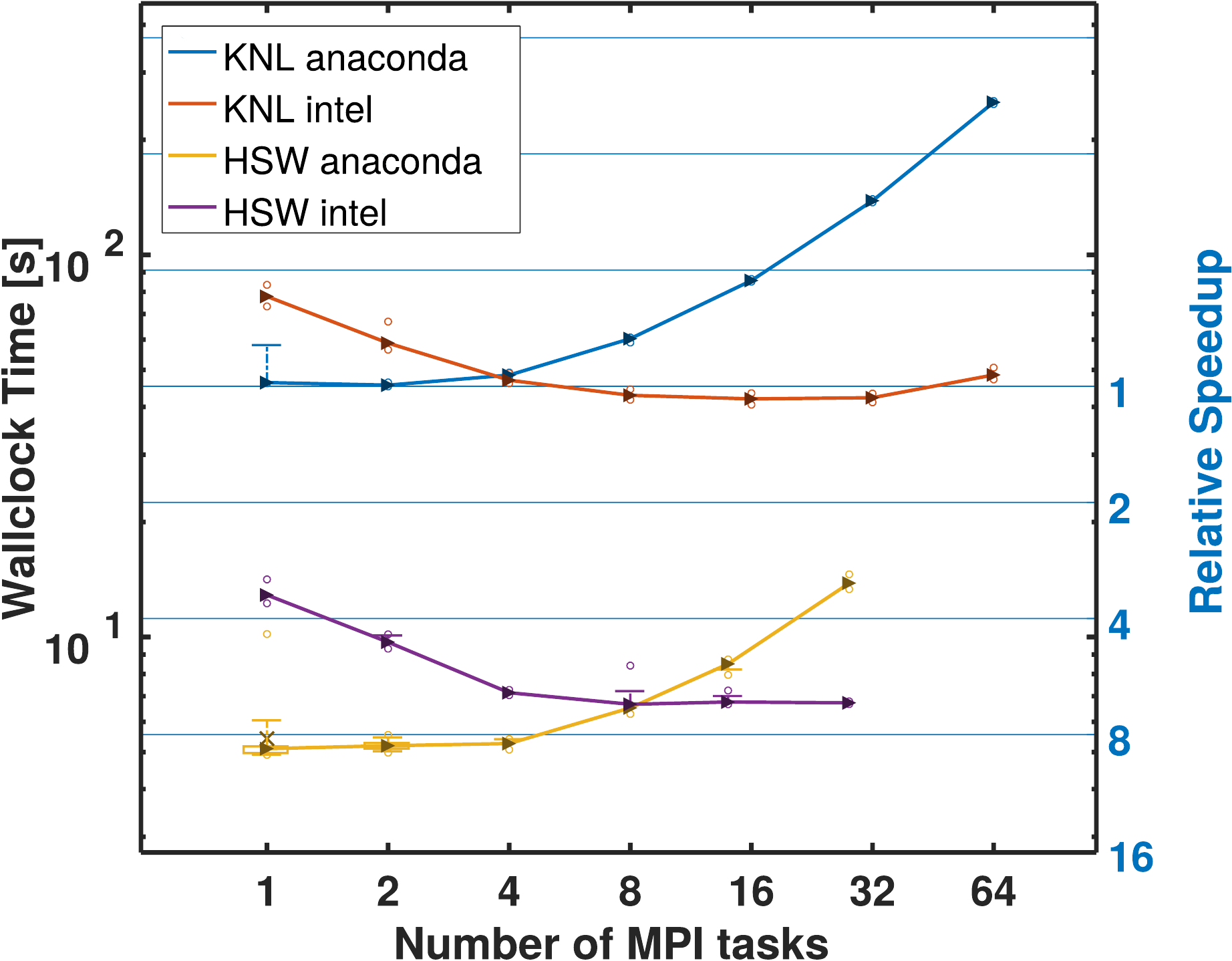}
    \caption{Scaling plots of \yt{} projections on single KNL and HSW nodes and for \Intel{} and Anaconda \py{} distribution, as indicated in the legend.}\label{f:yt_projection}
\end{figure}

In Figure \ref{f:yt_projection} we present scaling curves for both Anaconda and \Intel{} \py{}, on KNL and, for comparison, on HSW.
{The disappointing scaling shows clearly that many-core systems are not indicated for producing projections.} However we can draw a number of useful considerations:
\begin{itemize}
\item our measured scaling is at odds with the good 
{one} shown by \cite{turk2011yt} but 
{this tension can be explained when considering the different workloads.}
First, the dataset used in \cite{turk2011yt} has about a factor of 50 more grid elements 
than ours. 
{This increased workload can help the scaling; also, parallelizing across several nodes as they do could be more beneficial than using more cores on a single many-core node. This is because, as we measure, the memory overhead becomes more significant with more tasks, and adding more nodes brings additional memory capacity.}

\item Anaconda Python shows better performance in serial but also no 
parallel speedup (blue line in Figure \protect\ref{f:yt_projection}). Interestingly, \Intel{} \py{} has opposite trends: its time to solution is worse on few MPI tasks but then it slightly improves, without significant overhead at large number of tasks (red line in Figure \ref{f:yt_projection}).

\item Without a decent scaling, it is not surprising that the performance is much better on HSW than on KNL. However, while on HSW Anaconda performs best (yellow line in Figure \ref{f:yt_projection}), on KNL both distributions provide almost identical results, in terms of best performance on the node (i.e.~minimum on the scaling curve). This finding confirms that \Intel{} \py{}, despite the unsatisfactory performance of \yt{} in this specific problem, is successful in targeting many-core systems. 
{In forthcoming work we extend our investigation to other analysis types, testing the performance on the SKX architecture}  \cite{cielo2019ytpum}.
\end{itemize}

As already stated above, using \mpiforpy{} is not the only parallelization strategy available in \yt{}, therefore 
{we decided to implement these other solutions, and test them on KNL hardware as well. We present the results in the next Section.}

\subsection{\yt{}: volume rendering with \cy{}}\label{ss:yt-render}

We further set to explore the performance gain and the parallel capabilities achievable through \cy{}, when used -- in one of its simplest forms -- as an optimized \py{} compiler. Integration with \cy{} is encouraged and supported by \yt{}, and gives access to 
{additional} parallelization schemes. When used for volume rendering, the parallelism can be exploited with \mpiforpy{}, as we did in Section \ref{ss:yt-projection}, with OpenMP (automatic by \cy{} interface) or even with both, in hybrid mode. 
We thus setup a volume rendering test-script, using the same \enzo{} file as our \visit{} renderings (Section \ref{ss:visit-results}) and \yt{} density projections (Section \ref{ss:yt-projection}). We used a linear ramp transmission function, as instructed by the \yt{} documentation \cite{yt201Xrender}, but within a custom-defined function we could compile and build\footurl{http://docs.cython.org/en/latest/} with \cy{}.  This whole operation is very affine to what shown for \visit{} (top panel of the same figure), although now we use the full dataset and not just the central high-resolution region (still visible in the image center). The produced image is shown next to that rendering, in Figure \ref{f:visit_visual} (bottom panel).

\begin{figure}    
    \centering
    \includegraphics[width=\columnwidth]{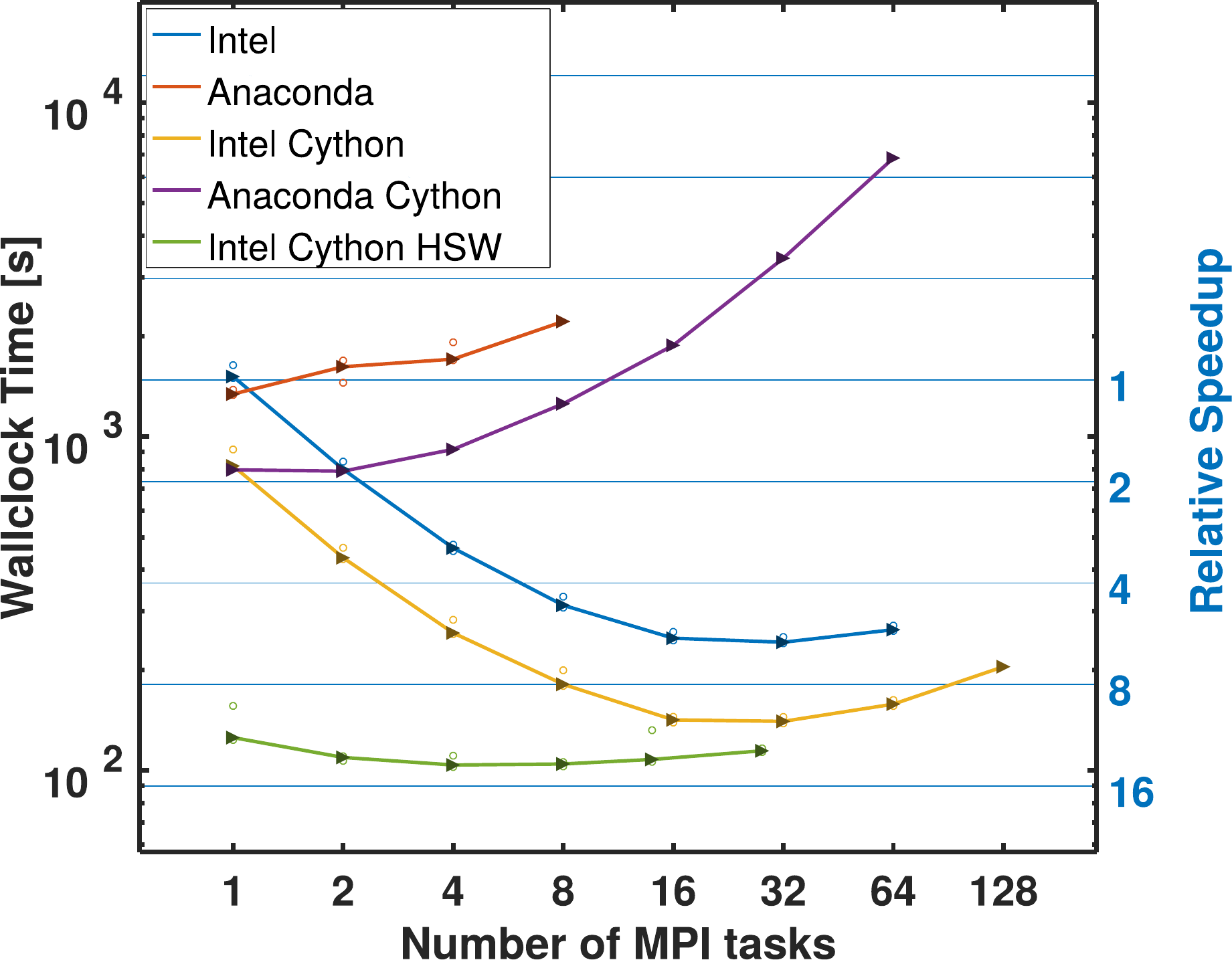}
\caption{Parallel scaling for 3D rendering with yt, using \cy{} and \mpiforpy{}.
The meaning of the symbols is the same as in the legend of Figure \ref{f:echo_scaling}.
\Intel{} \py{} shows scalability up to 16 cores; Anaconda never scales. 
\cy{} improves all performances by about $2\times$.
On HSW, the serial performance is better but the parallel speedup negligible.} \label{f:yt_rendering_mpi}
\end{figure}

The results of the scaling tests using pure \mpiforpy{} are displayed in Figure \ref{f:yt_rendering_mpi}. We show Anaconda and Intel \py{} (violet and yellow lines), together with the scaling obtained by both without using \cy{}, but only \mpiforpy{} (red and blue lines). Intel \py{} achieves moderate but satisfactory scaling, up to $6\times$ for 16 cores.
Hyperthreading is not beneficial, as it would require good scaling over the whole node in the first place.

For comparison, the standard Anaconda appears incapable of any scaling, with or without \cy{}.  The time to solution instead increases up to very large factors due to communication override. Past 4 or 8 processes also the memory footprint of the application grows, to occupy almost all the available DRAM. However this happens also with the \Intel{} version, thus it is not the reason behind the poor scaling. 
While larger workload sizes or different input formats may yield better results,
{this scaling behavior is seriously alarming, as the parallelization strategy involving Anaconda \py{} and \cy{} is suggested as the default in the \yt{} documentation, and most users may choose not to go beyond this scheme.}

The effect of \cy{} optimization is very beneficial, as it provides a 
speedup of about $2\times$ over any counterpart setup without it, leaving the scaling behavior essentially unaltered.

In the same figure we plot also the scaling of our best instance, \Intel{} with \cy{}, on a HSW node (green line). While the serial performance is still more than $6\times$ better on HSW, there is no speedup past 4 cores, so that the performance difference on the whole nodes tends to disappear, at variance with what observed with the projections in Section \ref{ss:yt-projection}, and similar to what seen for \gadget{} (Section \ref{s:gadget-perf}). This confirms that the improvements of \Intel{} \py{} are specific to architectures 
{that are} more recent and 
{expose more parallelism} than HSW. Somewhat sadly, using embarrassingly parallel schemes on HSW would still be the most performing, and least memory-consuming, option.

\begin{figure}    
    \centering
    \includegraphics[width=\columnwidth]{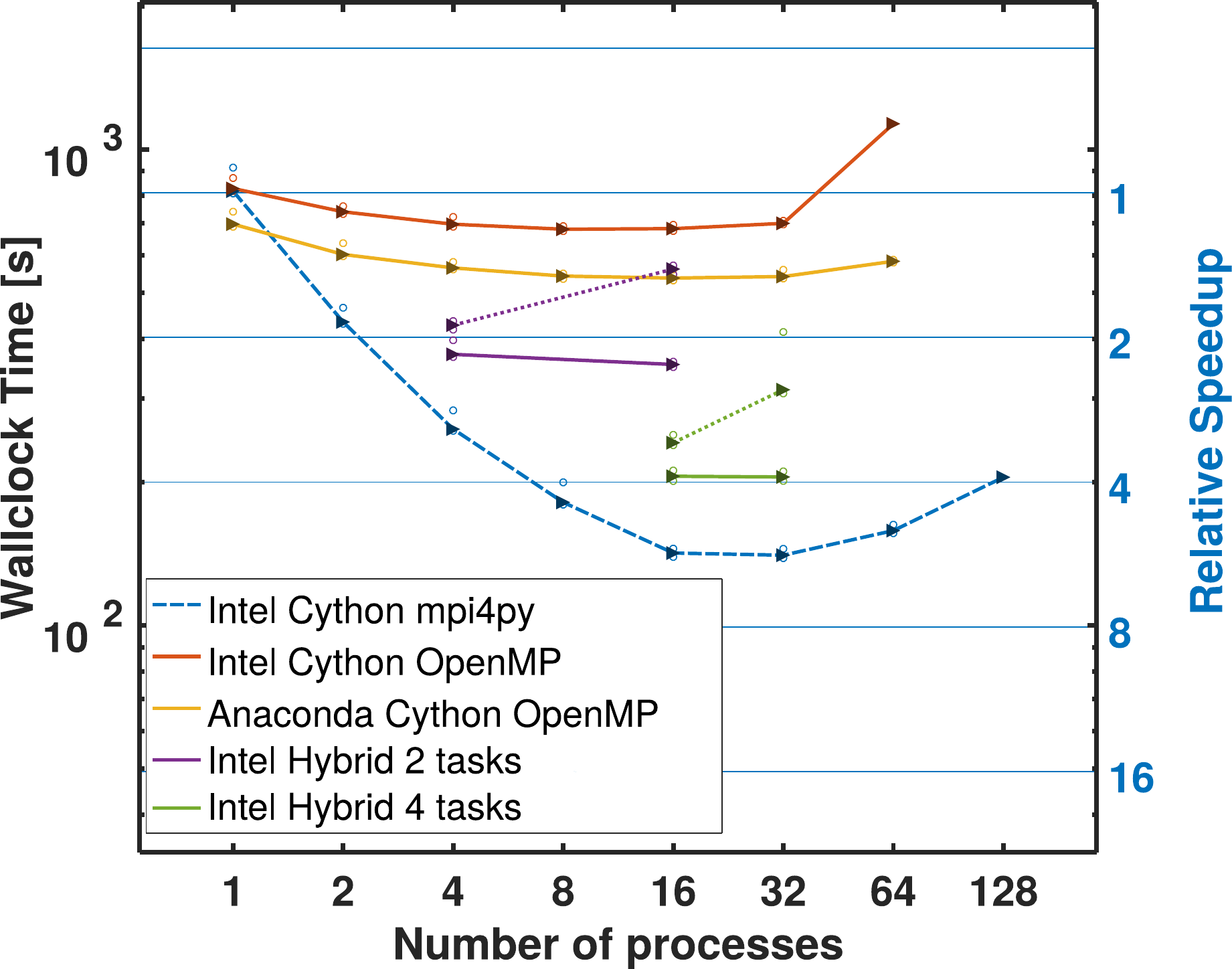}
\caption{Parallel scaling for 3D rendering with \yt{}, via the built-in OpenMP interface of \cy{}, used also in hybrid mode with 2 or 4 MPI tasks. The symbols have the same meaning as Figure \ref{f:echo_scaling}, and the \mpiforpy{} line of Figure \ref{f:yt_rendering_mpi} (dashed) is plotted again for reference. For hybrid mode, the results of varying process pinning are shown as well (solid: default, dotted: scatter). The scaling provided by OpenMP is measurable but negligible, always outperformed by the pure \mpiforpy{} scheme.}
    \label{f:yt_rendering_omp}
\end{figure}

In Figure \ref{f:yt_rendering_omp} we show the results of scaling tests for the same rendering task, using the automatic OpenMP interface provided by \cy, and a hybrid \mpiforpy{} + OpenMP mode. The best scaling from Figure \ref{f:yt_rendering_mpi} is plotted as well for reference (dashed line).
While the OpenMP setup required no additional work, it brings 
{almost} no benefit either.
The speedup is measurable (validating the correctness of our setup) but negligible, 
always way below $2\times$; the absolute performance in this case is best with Anaconda. The hybrid cases, with 2 or 4 \mpiforpy{} tasks, and up to 8 OpenMP threads are no better: at best, they show a very small speedup as for pure OpenMP, and they are always outperformed by the pure \mpiforpy{}. The solid lines refer to the default process pinning (i.e. \texttt{I\_MPI\_PIN\_DOMAIN=auto}, \texttt{KMP\_AFFINITY} unset), while the dotted lines refer to a scheme scattering the processes around the whole node (i.e. \texttt{I\_MPI\_PIN\_DOMAIN=auto}, \texttt{KMP\_AFFINITY=scatter}). The latter leads always to a performance degradation, in line with the general behavior observed for \yt{} on the KNL: the larger the fraction of the node used, the smaller the gain.
    
\section{Summary and conclusions}\label{s:conclusion}

{The aim of this work is to provide practical insights to HPC users interested in modernization of hardware and scientific code.}
We considered a pool of codes relevant for astrophysics, including diverse simulation codes (\gadget{}, 
{\flash{} and \echo{}}) and two among the most diffuse post-processing and visualization tools (\visit{} and \yt{}).
The parallel performance characterization of this pool is able to touch most of the relevant aspects for the field of HPC, in sight of the Exascale computing era. We conduct our performance investigation on KNL nodes, as their features embody many foreseen aspects of near-future, massively-parallel architectures.
{Existing KNL nodes and clusters are still profitably used for production in the field, however, here we highlight their value as code optimization laboratories, regardless of absolute performance. In other words, a code capable of \emph{ticking all the boxes} of the KNL, i.e., taking advantage of all its peculiar features, will perform very well also on \Intel{} \Xeon{} of current and previous generations. Astrophysical users whose codes fail to do so, may still find better performance on \Xeon{} products, even older ones, but need to pay more attention to optimization on (pre)exascale systems. For those users, this paper provides some useful real-world examples. On the other end, vendors interested in astrophysical applications should be aware of those boxes that no or few scientific codes are able to tick (e.g., the flat MCDRAM memory mode).}

We {are confident} that our results will provide useful guidelines for the design of new codes and the optimization of existing ones, as well as for the {co-}design of new hardware that may fit better the needs of a scientific community growing more and more aware of the possibilities and challenges presented by HPC.

{We start by discussing our results about simulation codes.}
Large parts of the simulation codes of our choice are developed by 
domain scientists, rather than by professional programmers, to solve a specific class of problems. 
{\flash{} stands in mid-ground, as its fundamental routines and data layout are developed and maintained by HPC professionals (and this is reflected in its base performance), while scientific users code their own setups, thus contributing to the enrichment of the physics libraries. }

{One important} downside 
{of user-driven development} is that most core aspects of these codes were not originally designed to achieve high parallel performance (data layout, memory and cache optimization, SIMD). In turn, this has lead to successive rounds of optimization work, such as the recent optimization histories we presented in this paper for \gadget{}, \flash{} and \echo{}, that have proven very beneficial across different  architectures, to the point that all these codes can efficiently be run on modern supercomputers, albeit each with its own peculiar strengths and weaknesses.

For the considered simulation codes, two main points emerge. The first is that the availability of a large number of cores is beneficial in all cases, for typical application workloads. 
{The parallel speedup achieved by optimized versions is very encouraging, both at the level of a single KNL node (\gadget{} and \echo{}, OpenMP-based) and across many nodes (\flash{})}.
In this respect the advancements in the parallelization paradigms (MPI, as seen with \flash{}, 
and OpenMP), as well as the efforts to introduce hybrid parallel schemes and improve memory access efficiency (\gadget{}, \echo{}) have been very valuable: optimization work targeted to previous generation architectures brings added benefit on many-core systems, 
{without any additional} work necessary. The second point is that in many cases the nature of the applications prevents them from taking more than modest advantage of other degrees of parallelism, such as vectorization 
{($\sim1.2\times$ speedup fot both \flash{} and \echo{})}
and SMT
{and features such as the \emph{flat} memory mode}. The 
{tested simulation codes} are 
{mostly bandwidth-bound}, calling for further dedicated optimization work at fundamental level, such as cache-blocking techniques and fundamental revisions of computational algorithms and data-layout (\gadget{}). 

The discussion is different for data analysis/visualization codes.
Traditionally, data post-processing in astrophysics is conducted {serially} or on few CPUs; visualization on one or few GPUs. However code scaling is increasingly important with increasing size and complexity of input files (more complex and larger simulations, but also observations in the \emph{Big Data} regime from the newest generation of telescopes). 
{At variance with the simulation codes, the}
data analysis applications we consider are typically built upon rather general-purpose, polished algorithms (e.g. raytracing), and HPC-aware building blocks (e.g. \py{}). 
{As a consequence, we found the codes more optimized to begin with, but they end up providing less freedom in performance analysis or tuning, as we summarize below.}

Concerning scaling, the scenario appears mixed.
{Codes such as \visit{} include algorithms explicitly designed for HPC systems, and thus}
are capable of efficient parallel scaling, making solutions like software-defined visualization very promising, also targeting hardware successive to the KNL (such as SKX). Thanks to its good scaling, \visit{} can also take good advantage of hyperthreading to fill very efficiently its pipeline schedule. 

{On the other hand \yt{} focuses }
less on HPC optimization and more on the synergies given by its comprehensive \py{} environment, that however still presents some undeniable bottlenecks in HPC.
Some of these issues are being addressed by tools like \Intel{} \py{} and \cy{}; we have shown how these increase respectively scalability (on KNL more than HSW) and performance over 
{the standard} Anaconda \py{} distribution. This is especially true for high-workload tasks such as volume rendering, though 
{the scaling does not allow the full utilization of the parallelism exposed by all cores in the node,} thus room for improvement persists.
It must be noted that \yt{}'s algorithm is rather performing in serial execution, and that embarassingly-parallel workload sharing is a viable solution for practical usage, although sub-optimal in HPC context. 
On this topic of algorithm performance, both \visit{} and \yt{} exemplify how having a pool of algorithms to choose from is a winning strategy, since the 
{most performing options} in serial are not necessarily the best for large systems, as in the case of the \kb{} sampling algorithm for ray-casting with \visit{}.
Likewise, the possibility of interchanging libraries and building blocks is a big advantage, as more and more HPC- and performance-focused packages are deployed (\mesa{}, \Intel{} \py{}, \cy{}, etc.), although the benefits largely vary with the considered application. 

{The  two  codes  present  some  common  limitations  as  well, mostly concerning hybrid parallelization and vectorization. However the most recent software stacks (e.g., \Intel{} MPI and \py{} 2019, or the aforementioned \ospray{}) can provide access to these features on both \Xeon{} and \XeonPhi{}, sometimes with minimal or no additional work for the user.}

Finally, to look at the same results above but from the hardware side, 
what are the architectural features of KNL that are most useful for the tested applications, and that look most promising if adopted in a prospective pre-Exascale system?

We noted already that the high number of available cores per node is a feature most codes take advantage of, even those with non-hybrid parallelization. From this perspective it is good news that this feature is carried along in \Xeon{} processors of the subsequent generation. 

Indeed, the second generation of the Intel\textsuperscript{\textregistered} Xeon\textsuperscript{\textregistered} Scalable Processor, codename CascadeLake, is available also as a $56$ cores CPU. In this case the multi-threading capability of the codes and the porting performance can be guaranteed as well. Such systems can be considered as a current, viable successor architecture of the KNL. One should however keep in mind that the use of accelerators is nowadays a crucial question, when planning an investment on hardware.

Some code may achieve an additional speedup with a 2 processes-per-core SMT, thus enabling hyper-threading up to {that level} is desirable, though the use of more than two threads per physical core seldom brings further performance improvement.
{The AVX512 instruction set was kept and improved in the following Xeon Scalable processors, together with version of MKL, TBB, \py{} and other libraries that include AVX512 versions, making them more easily accessible by high-level applications such as \visit{} and \yt{}. This is often the first step in the optimization of this class of codes.}

Other peculiar KNL features did not contribute as much as expected in terms of performance: the most striking example is MCDRAM. While in principle it addresses
a serious bottleneck of modern codes (many of which are bandwidth-limited),
the \emph{flat mode} with allocation on high-bandwidth memory does not really introduce significant performance difference
{(as summarized in Table \ref{t-kpi}))}. Moreover, for memory-intensive applications (as most of the astrophysical ones), the \emph{cache} mode is in fact the only viable choice. However, tradeoffs with a more traditional L3 cache (smaller, faster) are difficult to evaluate. It appears also that most of the tested codes require additional work to fully profit from large vector registers, as vectorization remains a nearly unused resource,
{with the exception of \flash{}}. This issue is coupled to the access of data into memory: if this is not cache-aware, it can spoil vectorization, even if it is allowed by the compiler.
{In general, the gap between memory and CPU performance (bandwidth and per-core capacity) is a particularily hard bottleneck for numerical science applications (as for instance shown in our roofline plot, in Sec. \ref{s:echo}). Fortunately, vendors such as Intel are aware of such issues, thus they keep developing advanced solutions throughout the whole memory hierarchy even after the MCDRAM (e.g., the} recent \Intel{}\textsuperscript{\textregistered}\ Optane\texttrademark\ DC memory \cite{optane2019intel}).

A positive note concerns the performance \emph{tuning} of the node. While scanning a very large parameter space (node configuration, compilation flags, ratio of MPI tasks over OpenMP threads) we have shown that in some cases one can get a moderate performance boost without any need of code development. Besides such tuning, we stress however that some degree of optimization is almost mandatory in practice. 

In conclusion, we have shown that the efficient utilization of modern HPC resources, to enable the next generation of scientific discovery on the road to Exascale, stems from the combination of suitable hardware features and applications which are able to exploit them. It would be extremely useful if future workflows can automate some basic operations (memory management via cache blocking, NUMA-aware memory management, process pinning, to name a few) which are crucial but currently are left to the developers. The portability of such solutions is also a concern.

\section*{Acknowledgment}\label{s:ack}
Part of this work was financially supported by the PRACE project funded in part by the EU’s Horizon 2020 Research and Innovation programme (2014-2020) under grant agreement 653838. 

The research of L.~I. has been partly supported by the Intel Parallel Computing Center (Intel PCC) ExScaMIC-KNL - Extreme Scaling on Intel Xeon Phi Supercomputers at LRZ and Technical University of Munich (TUM). 

M.~B. acknowledges support from the European Research Council (grant no. 715368 -- MagBURST) and from the Gauss Centre for Supercomputing e.V. ({\tt www.gauss-centre.eu}) for funding this project by providing computing time on the GCS Supercomputer SuperMUC at Leibniz Supercomputing Centre ({\tt www.lrz.de}). 

C.~F.~acknowledges funding provided by the Australian Research Council (Discovery Project DP170100603 and Future Fellowship FT180100495), and the Australia-Germany Joint Research Cooperation Scheme (UA-DAAD). We further acknowledge high-performance computing resources provided by the Leibniz Rechenzentrum and the Gauss Centre for Supercomputing (grants~pr32lo, pr48pi and GCS Large-scale project~10391), the Australian National Computational Infrastructure (grant~ek9) in the framework of the National Computational Merit Allocation Scheme and the ANU Merit Allocation Scheme. The simulation software FLASH was in part developed by the DOE-supported Flash Center for Computational Science at the University of Chicago.

Intel, Xeon and Xeon Phi are trademarks of Intel Corporation or its subsidiaries in the U.S. and/or other countries.


\bibliographystyle{IEEEtran.bst}
\bibliography{IEEEabrv,sample.bib}

\begin{thebibliography}{10}
\providecommand{\url}[1]{#1}
\csname url@samestyle\endcsname
\providecommand{\newblock}{\relax}
\providecommand{\bibinfo}[2]{#2}
\providecommand{\BIBentrySTDinterwordspacing}{\spaceskip=0pt\relax}
\providecommand{\BIBentryALTinterwordstretchfactor}{4}
\providecommand{\BIBentryALTinterwordspacing}{\spaceskip=\fontdimen2\font plus
\BIBentryALTinterwordstretchfactor\fontdimen3\font minus
  \fontdimen4\font\relax}
\providecommand{\BIBforeignlanguage}[2]{{%
\expandafter\ifx\csname l@#1\endcsname\relax
\typeout{** WARNING: IEEEtran.bst: No hyphenation pattern has been}%
\typeout{** loaded for the language `#1'. Using the pattern for}%
\typeout{** the default language instead.}%
\else
\language=\csname l@#1\endcsname
\fi
#2}}
\providecommand{\BIBdecl}{\relax}
\BIBdecl

\bibitem{ColindeVerdiere2015}
G.~J.~L. Colin~de Verdiere, ``{Computing element evolution towards Exascale and
  its impact on legacy simulation codes},'' \emph{Eur. Phys. J.}, vol. A51,
  no.~12, p. 163, 2015.

\bibitem{mlb16}
\BIBentryALTinterwordspacing
A.~{Mathuriya}, Y.~{Luo}, A.~{Benali}, L.~{Shulenburger}, and J.~{Kim},
  ``{Optimization and parallelization of B-spline based orbital evaluations in
  QMC on multi/many-core shared memory processors},'' \emph{arXiv e-prints},
  2016, 1611.02665. [Online]. Available:
  \url{http://adsabs.harvard.edu/abs/2016arXiv161102665M}
\BIBentrySTDinterwordspacing

\bibitem{mrk17}
\BIBentryALTinterwordspacing
P.~J. {Mendygral}, N.~{Radcliffe}, K.~{Kandalla}, D.~{Porter}, B.~J. {O'Neill},
  C.~{Nolting}, P.~{Edmon}, J.~M.~F. {Donnert}, and T.~W. {Jones}, ``{WOMBAT: A
  Scalable and High-performance Astrophysical Magnetohydrodynamics Code},''
  \emph{\apjs}, vol. 228, p.~23, 2017. [Online]. Available:
  \url{http://adsabs.harvard.edu/abs/2017ApJS..228...23M}
\BIBentrySTDinterwordspacing

\bibitem{wkq17}
\BIBentryALTinterwordspacing
J.~W. {Wadsley}, B.~W. {Keller}, and T.~R. {Quinn}, ``{Gasoline2: a modern
  smoothed particle hydrodynamics code},'' \emph{\mnras}, vol. 471, pp.
  2357--2369, 2017. [Online]. Available:
  \url{http://adsabs.harvard.edu/abs/2017MNRAS.471.2357W}
\BIBentrySTDinterwordspacing

\bibitem{sgc16}
\BIBentryALTinterwordspacing
M.~{Schaller}, P.~{Gonnet}, A.~B.~G. {Chalk}, and P.~W. {Draper}, ``{SWIFT:
  Using task-based parallelism, fully asynchronous communication, and graph
  partition-based domain decomposition for strong scaling on more than 100,000
  cores},'' \emph{arXiv e-prints}, 2016, 1606.02738. [Online]. Available:
  \url{http://adsabs.harvard.edu/abs/2016arXiv160602738S}
\BIBentrySTDinterwordspacing

\bibitem{ipcc17}
F.~{Baruffa}, L.~{Iapichino}, N.~J. {Hammer}, and V.~{Karakasis}, ``Performance
  optimisation of smoothed particle hydrodynamics algorithms for
  multi/many-core architectures,'' in \emph{2017 International Conference on
  High Performance Computing Simulation (HPCS)}, 2017, pp. 381--388.

\bibitem{pwt18}
\BIBentryALTinterwordspacing
D.~J. {Price}, J.~{Wurster}, T.~S. {Tricco}, C.~{Nixon}, S.~{Toupin},
  A.~{Pettitt}, C.~{Chan}, D.~{Mentiplay}, G.~{Laibe}, S.~{Glover}, C.~{Dobbs},
  R.~{Nealon}, D.~{Liptai}, H.~{Worpel}, C.~{Bonnerot}, G.~{Dipierro},
  G.~{Ballabio}, E.~{Ragusa}, C.~{Federrath}, R.~{Iaconi}, T.~{Reichardt},
  D.~{Forgan}, M.~{Hutchison}, T.~{Constantino}, B.~{Ayliffe}, K.~{Hirsh}, and
  G.~{Lodato}, ``{Phantom: A Smoothed Particle Hydrodynamics and
  Magnetohydrodynamics Code for Astrophysics},'' \emph{Proc. Astron. Soc.
  Aust.}, vol.~35, p. e031, 2018. [Online]. Available:
  \url{http://adsabs.harvard.edu/abs/2018PASA...35...31P}
\BIBentrySTDinterwordspacing

\bibitem{xeonphi2018discontinued}
\BIBentryALTinterwordspacing
``Product change notification 116378 - 00,'' Intel Corporation, Tech. Rep.,
  2018. [Online]. Available:
  \url{https://qdms.intel.com/dm/i.aspx/9C54A9A7-BF37-4496-B268-BD2746EA54D3/PCN116378-00.pdf}
\BIBentrySTDinterwordspacing

\bibitem{knl17}
\BIBentryALTinterwordspacing
V.~{Codreanu} and J.~{Rodriguez}, \emph{{Best Practice Guide Knights Landing}},
  PRACE, 2017. [Online]. Available:
  \url{http://www.prace-ri.eu/best-practice-guide-knights-landing-january-2017/}
\BIBentrySTDinterwordspacing

\bibitem{zhang16}
\BIBentryALTinterwordspacing
B.~{Zhang}, ``Guide to automatic vectorization with intel avx-512 instructions
  in knights landing processors,'' Colfax International, Tech. Rep., 2016.
  [Online]. Available: \url{https://colfaxresearch.com/knl-avx512/}
\BIBentrySTDinterwordspacing

\bibitem{asai16b}
\BIBentryALTinterwordspacing
R.~{Asai}, ``Mcdram as high-bandwidth memory (hbm) in knights landing
  processors: Developer’s guide,'' Colfax International, Tech. Rep., 2016.
  [Online]. Available: \url{https://colfaxresearch.com/knl-mcdram/}
\BIBentrySTDinterwordspacing

\bibitem{va16}
\BIBentryALTinterwordspacing
A.~{Vladimirov} and R.~{Asai}, ``Clustering modes in knights landing
  processors: Developer’s guide,'' Colfax International, Tech. Rep., 2016.
  [Online]. Available: \url{https://colfaxresearch.com/knl-numa/}
\BIBentrySTDinterwordspacing

\bibitem{springel2005cosmological}
V.~{Springel}, ``{The cosmological simulation code GADGET-2},'' \emph{\mnras},
  vol. 364, pp. 1105--1134, 2005.

\bibitem{bma2016sph}
A.~M. {Beck}, G.~{Murante}, A.~{Arth}, R.-S. {Remus}, A.~F. {Teklu}, J.~M.~F.
  {Donnert}, S.~{Planelles}, M.~C. {Beck}, P.~{F{\"o}rster}, M.~{Imgrund},
  K.~{Dolag}, and S.~{Borgani}, ``{An improved SPH scheme for cosmological
  simulations},'' \emph{\mnras}, vol. 455, pp. 2110--2130, 2016.

\bibitem{abb14}
A.~Auweter, A.~Bode, M.~Brehm, L.~Brochard, N.~Hammer, H.~Huber, R.~Panda,
  F.~Thomas, and T.~Wilde, ``A case study of energy aware scheduling on
  supermuc,'' in \emph{Supercomputing}, J.~M. Kunkel, T.~Ludwig, and H.~W.
  Meuer, Eds.\hskip 1em plus 0.5em minus 0.4em\relax Cham: Springer
  International Publishing, 2014, pp. 394--409.

\bibitem{esc17}
\BIBentryALTinterwordspacing
J.~{Eastep}, S.~{Sylvester}, C.~{Cantalupo}, B.~{Geltz}, F.~{Ardanaz},
  A.~{Al-Rawi}, K.~{Livingston}, F.~{Keceli}, M.~{Maiterth}, and S.~{Jana},
  \emph{{Global Extensible Open Power Manager: A Vehicle for HPC Community
  Collaboration on Co-Designed Energy Management Solutions}}.\hskip 1em plus
  0.5em minus 0.4em\relax Springer International Publishing, 2017, pp.
  394--412. [Online]. Available:
  \url{https://doi.org/10.1007/978-3-319-58667-0_21}
\BIBentrySTDinterwordspacing

\bibitem{sib19}
\BIBentryALTinterwordspacing
R.~{Sch{\"o}ne}, T.~{Ilsche}, M.~{Bielert}, A.~{Gocht}, and D.~{Hackenberg},
  ``{Energy Efficiency Features of the Intel Skylake-SP Processor and Their
  Impact on Performance},'' \emph{arXiv e-prints}, 2019, 1905.12468. [Online].
  Available: \url{https://ui.adsabs.harvard.edu/abs/2019arXiv190512468S}
\BIBentrySTDinterwordspacing

\bibitem{fryxell2000FLASH}
B.~Fryxell, K.~Olson, P.~Ricker, F.~Timmes, M.~Zingale, D.~Lamb, P.~MacNeice,
  R.~Rosner, J.~Truran, and H.~Tufo, ``{FLASH: An Adaptive Mesh Hydrodynamics
  Code for Modeling Astrophysical Thermonuclear Flashes},'' \emph{Astrophys. J.
  Supp.}, vol. 131, pp. 273--334, 2000.

\bibitem{fk12}
C.~{Federrath} and R.~S. {Klessen}, ``{The Star Formation Rate of Turbulent
  Magnetized Clouds: Comparing Theory, Simulations, and Observations},''
  \emph{\apj}, vol. 761, p. 156, Dec 2012.

\bibitem{fk13}
------, ``{On the Star Formation Efficiency of Turbulent Magnetized Clouds},''
  \emph{\apj}, vol. 763, p.~51, Jan. 2013.

\bibitem{antonuccio2010selfregulation}
V.~Antonuccio-Delogu and J.~Silk, ``{Active Galactic Nuclei Activity:
  Self-Regulation from Backflow},'' \emph{\mnras}, vol. 405, pp. 1303--1314,
  Jun. 2010.

\bibitem{cielo2018reorienting}
\BIBentryALTinterwordspacing
{Cielo, S.}, {Babul, A.}, {Antonuccio-Delogu, V.}, {Silk, J.}, and {Volonteri,
  M.}, ``Feedback from reorienting {AGN} jets - {I. Jet-ICM} coupling, cavity
  properties and global energetics,'' \emph{A\&A}, vol. 617, p. A58, 2018.
  [Online]. Available: \url{https://doi.org/10.1051/0004-6361/201832582}
\BIBentrySTDinterwordspacing

\bibitem{fki16}
C.~{Federrath}, R.~S. {Klessen}, L.~{Iapichino}, and N.~J. {Hammer}, ``{The
  world's largest turbulence simulations},'' \emph{arXiv e-prints}, p.
  arXiv:1607.00630, Jul 2016.

\bibitem{delzanna2007}
L.~Del~Zanna, O.~Zanotti, N.~Bucciantini, and P.~Londrillo, ``{ECHO}: a
  {Eulerian} conservative high-order scheme for general relativistic
  magnetohydrodynamics and magnetodynamics,'' \emph{\aap}, vol. 473, pp.
  11--30, 2007.

\bibitem{olmi2015}
\BIBentryALTinterwordspacing
B.~Olmi, L.~Del~Zanna, E.~Amato, and N.~Bucciantini, ``Constraints on particle
  acceleration sites in the {Crab} nebula from relativistic magnetohydrodynamic
  simulations,'' \emph{Monthly Notices of the Royal Astronomical Society}, vol.
  449, no.~3, pp. 3149--3159, May 2015. [Online]. Available:
  \url{https://academic.oup.com/mnras/article/449/3/3149/2892996}
\BIBentrySTDinterwordspacing

\bibitem{pili2014}
\BIBentryALTinterwordspacing
A.~G. Pili, N.~Bucciantini, and L.~Del~Zanna,
  ``\BIBforeignlanguage{en}{Axisymmetric equilibrium models for magnetized
  neutron stars in {General} {Relativity} under the {Conformally} {Flat}
  {Condition}},'' \emph{\BIBforeignlanguage{en}{Monthly Notices of the Royal
  Astronomical Society}}, vol. 439, no.~4, pp. 3541--3563, Apr. 2014. [Online].
  Available: \url{https://academic.oup.com/mnras/article/439/4/3541/1161070}
\BIBentrySTDinterwordspacing

\bibitem{pili2017}
\BIBentryALTinterwordspacing
------, ``General relativistic models for rotating magnetized neutron stars in
  conformally flat space–time,'' \emph{\mnras}, vol. 470, no.~2, pp.
  2469--2493, Sep. 2017. [Online]. Available:
  \url{https://academic.oup.com/mnras/article-lookup/doi/10.1093/mnras/stx1176}
\BIBentrySTDinterwordspacing

\bibitem{bugli2018a}
\BIBentryALTinterwordspacing
M.~Bugli, J.~Guilet, E.~M\"uller, L.~Del~Zanna, N.~Bucciantini, and P.~J.
  Montero, ``Papaloizou-{Pringle} instability suppression by the
  magnetorotational instability in relativistic accretion discs,''
  \emph{Monthly Notices of the Royal Astronomical Society}, vol. 475, pp.
  108--120, Mar. 2018. [Online]. Available:
  \url{http://adsabs.harvard.edu/abs/2018MNRAS.475..108B}
\BIBentrySTDinterwordspacing

\bibitem{bucciantini2013}
\BIBentryALTinterwordspacing
N.~Bucciantini and L.~Del~Zanna, ``\BIBforeignlanguage{en}{A fully covariant
  mean-field dynamo closure for numerical 3 + 1 resistive {GRMHD}},''
  \emph{\BIBforeignlanguage{en}{Monthly Notices of the Royal Astronomical
  Society}}, vol. 428, no.~1, pp. 71--85, Jan. 2013. [Online]. Available:
  \url{https://academic.oup.com/mnras/article/428/1/71/1059646}
\BIBentrySTDinterwordspacing

\bibitem{bugli2014}
\BIBentryALTinterwordspacing
M.~Bugli, L.~Del~Zanna, and N.~Bucciantini, ``\BIBforeignlanguage{en}{Dynamo
  action in thick discs around {Kerr} black holes: high-order resistive {GRMHD}
  simulations},'' \emph{\BIBforeignlanguage{en}{Monthly Notices of the Royal
  Astronomical Society: Letters}}, vol. 440, no.~1, pp. L41--L45, May 2014.
  [Online]. Available:
  \url{https://academic.oup.com/mnrasl/article/440/1/L41/1393473}
\BIBentrySTDinterwordspacing

\bibitem{delzanna2016}
L.~Del~Zanna, E.~Papini, S.~Landi, M.~Bugli, and N.~Bucciantini, ``Fast
  reconnection in relativistic plasmas: the magnetohydrodynamics tearing
  instability revisited,'' \emph{{\textbackslash}mnras}, vol. 460, pp.
  3753--3765, Aug. 2016.

\bibitem{bugli2018b}
\BIBentryALTinterwordspacing
M.~Bugli, ``{ECHO}-3dhpc: {Relativistic} {Accretion} {Disks} onto {Black}
  {Holes},'' in \emph{2018 26th {Euromicro} {International} {Conference} on
  {Parallel}, {Distributed} and {Network}-based {Processing} ({PDP})}, Mar.
  2018, pp. 674--681. [Online]. Available:
  \url{doi.ieeecomputersociety.org/10.1109/PDP2018.2018.00112}
\BIBentrySTDinterwordspacing

\bibitem{bugli2018c}
\BIBentryALTinterwordspacing
M.~Bugli, L.~Iapichino, and F.~Baruffa, ``Advancing the performance of
  astrophysics simulations with echo-3dhpc,'' \emph{\textcopyright Intel
  Parallel Universe Magazine}, vol.~34, pp. 49--56, Oct. 2018. [Online].
  Available:
  \url{https://software.intel.com/en-us/download/parallel-universe-magazine-issue-34-october-2018}
\BIBentrySTDinterwordspacing

\bibitem{resch2017}
\BIBentryALTinterwordspacing
{Resch, Michael}, ``{InSiDE} {Magazine} - {Innovative} {Supercomputing} in
  {Deutschland},'' Sep. 2017. [Online]. Available:
  \url{http://inside.hlrs.de/editions/17autumn.html}
\BIBentrySTDinterwordspacing

\bibitem{wwp09}
\BIBentryALTinterwordspacing
S.~Williams, A.~Waterman, and D.~Patterson, ``Roofline: An insightful visual
  performance model for multicore architectures,'' \emph{Commun. ACM}, vol.~52,
  no.~4, pp. 65--76, Apr. 2009. [Online]. Available:
  \url{http://doi.acm.org/10.1145/1498765.1498785}
\BIBentrySTDinterwordspacing

\bibitem{ips14}
A.~{Ilic}, F.~{Pratas}, and L.~{Sousa}, ``{Cache-aware Roofline model:
  Upgrading the loft},'' \emph{IEEE Computer Architecture Letters}, vol.~13,
  no.~1, pp. 21--24, 2014.

\bibitem{childs2012visit}
H.~Childs, E.~Brugger, B.~Whitlock, J.~Meredith, S.~Ahern, D.~Pugmire,
  K.~Biagas, M.~Miller, C.~Harrison, G.~H. Weber, H.~Krishnan, T.~Fogal,
  A.~Sanderson, C.~Garth, E.~W. Bethel, D.~Camp, O.~R\"{u}bel, M.~Durant, J.~M.
  Favre, and P.~Navr\'{a}til, ``{VisIt: An End-User Tool For Visualizing and
  Analyzing Very Large Data},'' in \emph{{High Performance
  Visualization--Enabling Extreme-Scale Scientific Insight}}, Oct 2012, pp.
  357--372.

\bibitem{wald2017ospray}
I.~Wald, G.~Johnson, J.~Amstutz, C.~Brownlee, A.~Knoll, J.~Jeffers,
  J.~Günther, and P.~Navratil, ``Ospray - a cpu ray tracing framework for
  scientific visualization,'' \emph{IEEE Transactions on Visualization and
  Computer Graphics}, vol.~23, no.~1, pp. 931--940, Jan 2017.

\bibitem{childs2006raycasting}
H.~Childs, M.~A.~Duchaineau, and K.-L. Ma, ``A scalable, hybrid scheme for
  volume rendering massive data sets .'' 01 2006, pp. 153--161.

\bibitem{ifk17}
\BIBentryALTinterwordspacing
L.~{Iapichino}, C.~{Federrath}, and R.~S. {Klessen}, ``{Adaptive mesh
  refinement simulations of a galaxy cluster merger - I. Resolving and
  modelling the turbulent flow in the cluster outskirts},'' \emph{\mnras}, vol.
  469, pp. 3641--3655, 2017. [Online]. Available:
  \url{https://ui.adsabs.harvard.edu/\#abs/2017MNRAS.469.3641I}
\BIBentrySTDinterwordspacing

\bibitem{enzo14}
\BIBentryALTinterwordspacing
G.~L. {Bryan}, M.~L. {Norman}, B.~W. {O'Shea}, T.~{Abel}, J.~H. {Wise}, M.~J.
  {Turk}, D.~R. {Reynolds}, D.~C. {Collins}, P.~{Wang}, S.~W. {Skillman},
  B.~{Smith}, R.~P. {Harkness}, J.~{Bordner}, J.-h. {Kim}, M.~{Kuhlen},
  H.~{Xu}, N.~{Goldbaum}, C.~{Hummels}, A.~G. {Kritsuk}, E.~{Tasker},
  S.~{Skory}, C.~M. {Simpson}, O.~{Hahn}, J.~S. {Oishi}, G.~C. {So}, F.~{Zhao},
  R.~{Cen}, Y.~{Li}, and {The Enzo Collaboration}, ``{ENZO: An Adaptive Mesh
  Refinement Code for Astrophysics},'' \emph{\apjs}, vol. 211, p.~19, 2014.
  [Online]. Available: \url{http://adsabs.harvard.edu/abs/2014ApJS..211...19B}
\BIBentrySTDinterwordspacing

\bibitem{atipa201Xospray}
\BIBentryALTinterwordspacing
``High performance computing and visualization - ground-breaking unified
  platform,'' Atipa Technologies, Tech. Rep. [Online]. Available:
  \url{https://www.atipa.com/hpc-visualization}
\BIBentrySTDinterwordspacing

\bibitem{wu2018visitospray}
Q.~Wu, W.~Usher, S.~Petruzza, S.~Kumar, F.~Wang, I.~Wald, V.~Pascucci, and
  C.~D. Hansen, ``{VisIt}-{OSPRay}: {Toward} an {Exascale} {Volume}
  {Visualization} {System},'' in \emph{Eurographics Symposium on Parallel
  Graphics and Visualization}, H.~Childs and F.~Cucchietti, Eds.\hskip 1em plus
  0.5em minus 0.4em\relax The Eurographics Association, 2018.

\bibitem{cielo2019supercomputing}
S.~Cielo, L.~Iapichino, G.~J., C.~Federrath, E.~Mayer, and M.~Wiedemann,
  ``{Visualizing the world’s largest turbulence simulation},'' \emph{Parallel
  Computing}, vol. SC19 Virtual Issue, 2019, in preparation.

\bibitem{turk2011yt}
M.~J. {Turk}, B.~D. {Smith}, J.~S. {Oishi}, S.~{Skory}, S.~W. {Skillman},
  T.~{Abel}, and M.~L. {Norman}, ``{yt: A Multi-code Analysis Toolkit for
  Astrophysical Simulation Data},'' \emph{\apjs}, vol. 192, p.~9, Jan. 2011.

\bibitem{intel201Xpythonbenchmarks}
\BIBentryALTinterwordspacing
``Intel distribution for \py{}, benchmarks - built for speed and scalability,''
  Intel Software, Tech. Rep. [Online]. Available:
  \url{https://software.intel.com/en-us/distribution-for-python/benchmarks}
\BIBentrySTDinterwordspacing

\bibitem{behnel2010cython}
S.~Behnel, R.~Bradshaw, C.~Citro, L.~Dalcin, D.~Seljebotn, and K.~Smith,
  ``Cython: The best of both worlds,'' \emph{Computing in Science Engineering},
  vol.~13, no.~2, pp. 31 --39, march-april 2011.

\bibitem{yt201Xparallel}
\BIBentryALTinterwordspacing
``Parallel computation with \yt{},'' the \yt{} project, Tech. Rep., 2013-2018.
  [Online]. Available:
  \url{https://yt-project.org/doc/analyzing/parallel_computation.html#parallel-computation-with-yt}
\BIBentrySTDinterwordspacing

\bibitem{cielo2019ytpum}
S.~Cielo, L.~Iapichino, and F.~Baruffa, ``{Speeding simulation analysis up with
  yt and Intel Distribution for Python},'' \emph{Intel Parallel Universe
  Magazine}, vol.~38, pp. 27--32, 2019.

\bibitem{yt201Xrender}
\BIBentryALTinterwordspacing
``3d visualization and volume rendering with \yt{},'' the \yt{} project, Tech.
  Rep., 2013-2018. [Online]. Available:
  \url{https://yt-project.org/doc/visualizing/volume_rendering.html?highlight=parallel#map-to-colormap}
\BIBentrySTDinterwordspacing

\bibitem{optane2019intel}
\BIBentryALTinterwordspacing
``Intel optane memory h10 with solid state storage product brief,'' Intel
  Corporation, Tech. Rep., 2019. [Online]. Available:
  \url{https://www.intel.com/content/dam/www/public/us/en/documents/product-briefs/optane-memory-h10-solid-state-storage-brief.pdf}
\BIBentrySTDinterwordspacing

\end{thebibliography}
\end{document}